\documentclass[lettersize,journal]{IEEEtran}

\usepackage{amssymb}
\usepackage{amsfonts}
\usepackage[fleqn]{amsmath} 
\usepackage{cite}
\usepackage{graphicx}
\usepackage{subfig}
\usepackage{epsf} 
\usepackage{pdfpages}
\usepackage{booktabs}
\usepackage{array}
\usepackage[font=normalsize,labelfont=bf, skip=20pt]{caption}
\usepackage{verbatim}
\usepackage{mathrsfs}
\usepackage{amssymb}
\usepackage{mathrsfs}
\usepackage{dsfont}
\usepackage[numbers, sort]{natbib}
\usepackage[ruled,linesnumbered]{algorithm2e}
\usepackage{algpseudocode}
\usepackage{xspace}
\usepackage[colorinlistoftodos]{todonotes}

\usepackage[normalem]{ulem}

\usepackage{mathtools, nccmath}

\usepackage{tikz}
\usepackage{textcomp}
\usepackage{titlesec}

\usepackage{wrapfig}
\usepackage{graphicx}

\usetikzlibrary{positioning, shapes.geometric}

\usepackage{setspace}
\usepackage{multirow}
\usepackage{url}

\ifCLASSINFOpdf
 
\else
 
\fi

\SetKwProg{Initialize}{Initialize:}{}{}

\usepackage{booktabs}
\usepackage{pgfplots}
\usepackage{pgfplotstable} 
\usepackage[hidelinks]{hyperref}

\newcommand{\etal}{\textit{et al.}\xspace}

\newtheorem{problem}{\textbf{Problem}}
\newtheorem{theorem}{Theorem}
\newtheorem{proof}{Proof}

\usepackage{atbegshi}
\AtBeginDocument{\AtBeginShipoutNext{\AtBeginShipoutDiscard}}

\begin{document}

\newcommand{\pt}[1]{\todo[author=pt]{#1}}

\thanks{This work has been submitted to the IEEE for possible publication. Copyright may be transferred without notice, after which this version may no longer be accessible.}

\title{Cross-Edge Orchestration of Serverless Functions with Probabilistic Caching}

\author{\IEEEauthorblockN{Chen Chen\IEEEauthorrefmark{1},
		Manuel Herrera\IEEEauthorrefmark{1}, Ge Zheng\IEEEauthorrefmark{1},
        Liqiao Xia\IEEEauthorrefmark{2}\IEEEauthorrefmark{1},
		Zhengyang Ling\IEEEauthorrefmark{1},  Jiangtao Wang\IEEEauthorrefmark{3}}\\
	\IEEEauthorblockA{
		\IEEEauthorrefmark{1}Department of Engineering,
		 University of Cambridge, UK \\
		\IEEEauthorrefmark{2}Department of Industrial and Systems Engineering, Hong Kong Polytechnic University, China\\
            \IEEEauthorrefmark{3}Research Centre for Intelligent Healthcare, Coventry University, UK\\
	}}

\maketitle

\pagestyle{empty}  
\thispagestyle{empty} 

\begin{abstract}\label{section:abstract}
Serverless edge computing adopts an event-based paradigm that provides back-end services on an as-used basis, resulting in efficient resource utilization.
To improve the end-to-end latency and revenue, service providers need to optimize the number and placement of serverless containers while considering the system cost incurred by the provisioning. 
The particular reason for this circumstance is that frequently creating and destroying containers not only increases the system cost but also degrades the time responsiveness due to the cold-start process.
Function caching is a common approach to mitigate the cold-start issue.
However, function caching requires extra hardware resources and hence incurs extra system costs.
Furthermore, the dynamic and bursty nature of serverless invocations remains an under-explored area.
Hence, it is vitally important for service providers to conduct a context-aware request distribution and container caching policy for serverless edge computing.
In this paper, we study the request distribution and container caching problem in serverless edge computing.
We prove the proposed problem is NP-hard and hence difficult to find a global optimal solution.
We jointly consider the distributed and resource-constrained nature of edge computing and propose an optimized request distribution algorithm that adapts to the dynamics of serverless invocations with a theoretical performance guarantee.
Also, we propose a context-aware probabilistic caching policy that incorporates a number of characteristics of serverless invocations. 
Via simulation and implementation results, we demonstrate the superiority of the proposed algorithm by outperforming existing caching policies in terms of the overall system cost and cold-start frequency by up to 62.1$\%$ and 69.1$\%$, respectively.
\end{abstract}



\IEEEpeerreviewmaketitle
\section{Introduction}\label{section:introduction}

With the accelerated penetration of 5G communications and Internet-of-Things (IoTs), a wide range of mobile and IoT applications are expected to be connected to the Internet, ranging from smart factories to edge computing-assisted online gaming, augmented reality and virtual reality \cite{Paraskevoulakou2023}, \cite{Shao2023}. 
As a result of deploying these diverse applications, a large amount of multi-modal data (e.g., video, image and audio) of the physical environment is continuously collected at various edge devices \cite{Xu2023}, \cite{Satapathy2023}.
To process such a tremendous amount of data in a time-responsive manner, edge computing has been introduced as a promising approach.
Edge computing, with a distributed nature, pushes computing resources and services from the network core to the network edges that are located in the proximity of mobile users and IoT applications, resulting in a significant reduction of end-to-end latency \cite{Mittal2021}, \cite{Xie2021}.

Recently, serverless computing has introduced new inspirations to edge computing and receives significant attention \cite{Pan2022}, \cite{Qi2022}.
Serverless computing is a service paradigm that provides simplified deployment, elastic computing, and fast responsiveness in an event-driven model \cite{Russo2023}, \cite{Stojkovic2023}.
Serverless paradigm encapsulates IoT services in lightweight containers and only deploys them when invoked by events requested by services \cite{Yu2023}, \cite{Deng2022}.
The containers are destroyed after completing the tasks and the hardware resources are released, providing resource efficiency and fast response simultaneously \cite{Kounev2023}.
For example, Microsoft Azure provides a comprehensive family of most advanced AI services, such as GPT-3.5, ChatGPT, DALL$\cdot$E, to enable intelligent, cutting-edge and responsible applications \cite{Ai57:online}.

\begin{figure}[t]
    \includegraphics[width=0.45\textwidth]{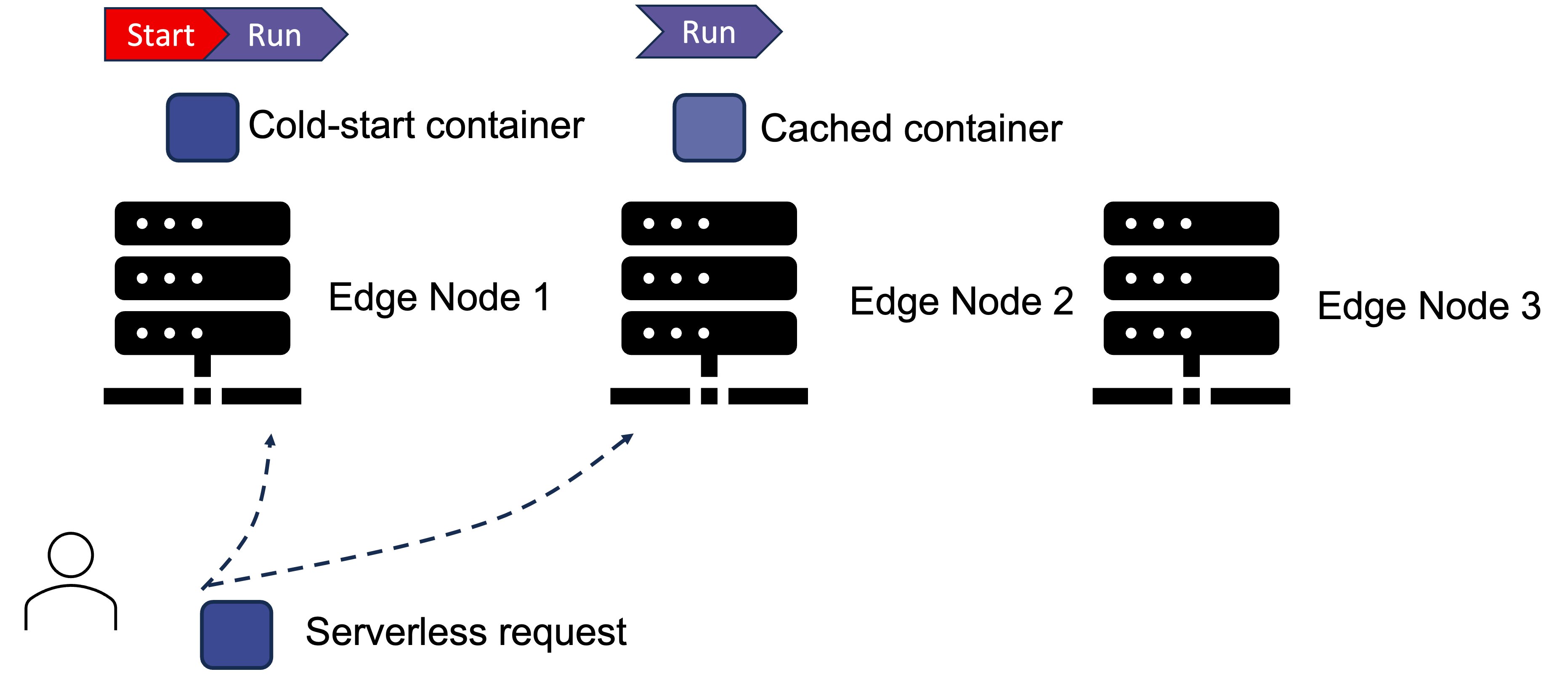}
    \centering
    \caption{Function caching in serverless edge computing}
    \label{fig:motivation}
\end{figure}

As illustrated in Figure~\ref{fig:motivation}, when end-user requests a serverless function, 
distributing containers near the data origin (edge node 1)
may lead to significant delay due to the cold-start process of creating a container \cite{Li2023}, \cite{Stojkovic2023b}, \cite{Schall2022}, \cite{Gu2022}.
Otherwise, the system can distribute the request to edge node 2, where a cached container may exist, to avoid cold-start issue at the cost of larger transmission delays.
Hence, unlike centralized cloud datacenters with homogeneous computing and storage substrates, supporting serverless funcitons between distributed edge nodes may offset the benefits of serverless paradigm.

When deploying various serverless containers across multiple geographically distributed edge nodes, fully optimizing the benefits of serverless containers is very challenging for the following reasons.
First, pioneering work largely overlooks the heterogeneous network topology in edge computing.
For instance, the origin of end-users may be distant from an edge node that holds the corresponding container.
Thus, besides container scheduling, how to jointly incorporate the data origin and containers for low-transmission delay is non-trivial for an efficient serverless platform \cite{Gu2023}.
Second, compared to cloud computing with seemingly endless computing resources, edge devices may regularly run at close to capacity \cite{Shang2023}, \cite{Cicconetti2021} and memory resources are still expensive at edge clouds \cite{Qi2022}, \cite{Li2023a}. 
Frequently creating and destroying containers significantly increases the system cost because creating containers requires downloading the code and starting a new execution environment before the requests can be served \cite{Liu2023}, \cite{Gu2021}.
Hence, optimizing container orchestration without statistical knowledge of future invocations is challenging.
Although function caching can be used to mitigate the above issue, keeping a function alive is not free.
Due to the pay-as-you-go model in serverless computing, users are not billed for functions that are not executing and hence the service providers need to undertake the cost of container caching. 
Hence, caching a number of containers in edge clouds massively increases the overall system cost.
Function caching will deteriorate service providers' revenue in the long term.
Therefore, balancing the container switching, communication and caching cost is pivotal to achieving an efficient system in edge clouds and this trade-off has not been explicitly studied so far.

Keep the above factors in mind, in this work, 
we aim to optimize the system efficiency defined by the combination of service latency and resource utilization.
We propose an online probabilistic caching policy to orchestrate serverless containers in edge computing, optimizing the system cost of the cross-edge serverless system. The proposed framework considers (1) the container switching cost that is proportional to the cold-start delay. (2) the communication cost incurred by the latency between edge nodes.
(3) the container running cost incurred by using hardware resources.
With the above setup, the cost minimization problem for cross-edge networks over the long term is formulated as integer linear programming (ILP) problem.

To optimize the unified cost, the proposed framework applies two policies: (1) A request distribution algorithm that dynamically assigns serverless requests to different edge nodes. This approach jointly considers the communication cost and the switching cost. (2) A context-aware probabilistic caching policy that captures the characteristics of serverless invocations. By doing so, it aims to improve the efficiency of caching in serverless edge computing environments.
Our main contributions are:
\begin{itemize}
\item We formulate a request distribution and container caching problem as an Integer Linear Programming (ILP) problem that jointly considers communication cost, container switching cost, and container running cost. 
We prove the NP-hardness of the problem, showing the complexity of finding a global optimal solution.
Then, we propose an online competitive algorithm for request distribution with a theoretical performance guarantee.

\item  We propose an online request distribution algorithm that adapts to the dynamics of incoming requests at each edge cloud. Due to the burstiness of serverless edge computing, workloads can vary significantly over time, and the ability to adjust caching strategies in response to these dynamics is pivotal. 

\item We propose a probabilistic caching policy (\textbf{pCache}) that captures the context of serverless functions, including the invocation frequency, memory usage and recency.
This approach is novel because it goes beyond simple caching strategies and considers the specific characteristics of each invocation in making caching decisions.
The awareness of serverless context sets the paper apart from existing caching policies.

\item To evaluate the performance, we use real-world traces and topology. We conduct extensive experiments over simulations and a serverless platform Knative \cite{HomeKnat85:online}, showing the superiority of our algorithm compared with state-of-the-art caching policies.
\end{itemize}

The remainder of the paper is organized as follows. 
In Section~\ref{section:relatedwork} we overview the related works of serverless edge computing, including request distribution and caching.
The state-of-the-art studies are summarized and the existing challenges are identified.
Section~\ref{section:system} introduces the system model and the system design. In Section~\ref{section:problem}, we formulate the optimization problem and prove its NP-hardness. Section~\ref{section:solution} presents our proposed algorithm, including the request distribution and the caching algorithms.
We also prove the theoretical bound of the proposed algorithms, proving the efficiency of the proposed algorithms.
Section~\ref{section:performance} evaluates the algorithm and conclusions are drawn in Section~\ref{section:conclusion}.
Moreover, we describe the prospects of serverless edge computing from our point of view and come up with some research opportunities.

\section{Related Work}\label{section:relatedwork}

A number of works investigate the function distribution problem in edge computing.
 Defuse \cite{Cao2018} addresses the issue of performance degradation caused by cold-starts in serverless platforms. 
  By constructing a function dependency graph, Defuse reduces the occurrences of cold-starts and demonstrates improved memory usage and decreased cold-start frequency compared to existing methods.
Chen \etal \cite{Chen2023} investigate the problem of container placement with latency optimization in edge computing environments. A priority-based algorithm is proposed to determine container termination in serverless computing.
The experimental results demonstrate that the proposed algorithm improves end-to-end response time.
Aslanpour \etal \cite{Aslanpour2022} focus on energy-aware scheduling in serverless edge computing. 
Zone-oriented and priority-based algorithms are proposed to improve the operational availability of bottleneck nodes, introducing concepts such as ``sticky offloading" and ``warm scheduling" for QoS optimization. 
Many other works \cite{Shen2021}, \cite{Fuerst2021}, \cite{Mittal2021} explore the cold-start problem in serverless computing but overlook the characteristics of serverless invocations. 

A wide range of research efforts have been carried out, focusing on service placement in edge computing. 
Poularakis \etal \cite{Poularakis2019} propose a joint optimization problem of service placement and request routing in Mobile Edge Computing (MEC) networks with multidimensional constraints. The proposed algorithm achieves near-optimal performance by using randomized rounding. 
Gu \etal \cite{Gu2022} focus on optimizing container-based microservice placement and request scheduling in edge computing. 
The potential of layer sharing among co-located microservices is adopted to enhance throughput and increase the number of hosted microservices. An iterative greedy algorithm is proposed with a guaranteed approximation ratio. 

\begin{figure*}[t]
    \includegraphics[width=0.8\textwidth]{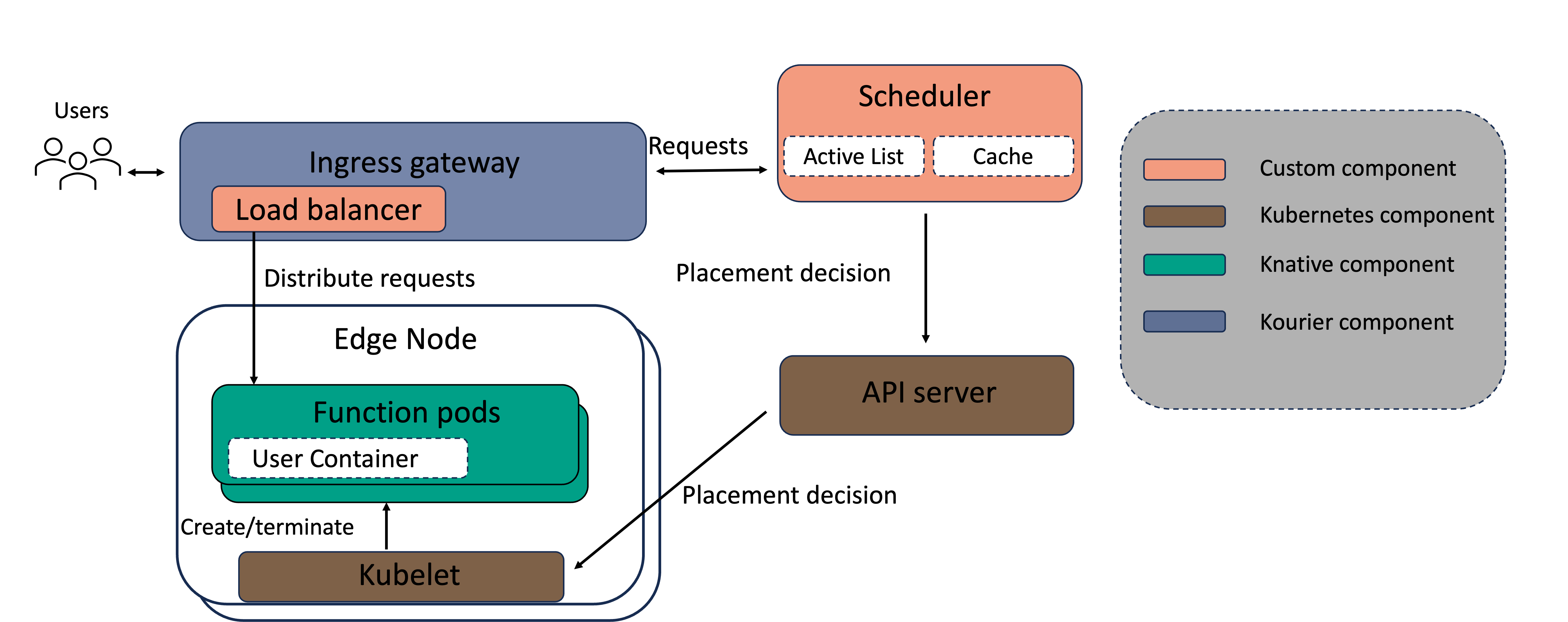}
    \centering
    \caption{pCache overview}
    \label{fig:system}
\end{figure*}

A few works also investigate the container scheduling problem.
Lin \etal \cite{Lin2021} investigate the challenges faced by cloud users when migrating applications in serverless computing. A heuristic algorithm called Probability Refined Critical Path Greedy algorithm (PRCP) is introduced to optimize the cost of serverless applications. The proposed models and algorithm are extensively evaluated on AWS Lambda and Step Functions, demonstrating their effectiveness.
Rausch \etal \cite{Rausch2021} introduce a container scheduling system called Skippy that optimizes task placement on edge infrastructures by considering factors such as data and computation movement, GPU acceleration, and operational objectives. The results demonstrate the effectiveness of Skippy in improving task placement quality and achieving operational goals.
Lin \etal \cite{Lin2023a} propose performance modeling and optimization algorithms by considering the cost in monetary terms.
The proposed algorithms provide prediction of performance and cost, helping developers to make decisions. 
These research efforts do not consider the dynamic and bursty nature of serverless invocations and hence are not applicable to our proposed problem.

In the context of edge computing, content caching is widely used as it can significantly reduce the delay in content delivery.
Tadrous \etal \cite{Tadrous2016} formulate a problem of proactive caching in mobile data networks, aiming to minimize service costs and improve delivery efficiency. 
The proposed solutions show close-to-optimal performance even with small prediction windows.
Jia \etal \cite{Jia2021} formulate a reliability-aware Service Function Chain (SFC) scheduling problem in a 5G network environment. A mixed integer non-linear programming problem is formulated. Also, an efficient algorithm for determining VNF redundancy and a reinforcement learning-based scheduling algorithm are proposed. Simulation results demonstrate the effectiveness of the proposed approach in increasing the success rate of dynamically arriving SFCs. 
Pan \etal \cite{Pan2022} investigate the retention-aware container caching problem in serverless edge computing. An optimization approach is proposed by using container caching and request distribution to improve system efficiency. By mapping the problem to the ski-rental problem and developing online algorithms, the study demonstrates significant performance gains compared to existing caching strategies. 
Farhadi \etal \cite{Farhadi2021} study the optimization of service placement and request scheduling in mobile edge computing environments. A two-time-scale framework is presented by jointly considering storage, communication, computation, and budget constraints. Extensive simulations demonstrate that the algorithm achieves near-optimal performance. 
Stephen \etal \cite{Pasteris2019} formulate a placement problem in heterogeneous mobile edge computing. The problem is proven to be NP-hard and a deterministic approximation algorithm is proposed with performance guarantee. The algorithm utilizes novel slot constructions on edge nodes and applies the method of conditional expectations for approximation guarantees. Simulation results demonstrate the superiority of the proposed algorithm over existing approaches. 
Zhou \etal \cite{Zhou2022} aim to optimize the service latency in UAV-assisted wirless mobile networks by incorporating the unique features of UAV.
To reduce the caching overhead, Lyapunov optimization approach and dependent rounding technique are adopted to achieve a near-optimal performance.
Cao \etal \cite{Cao2018} propose an optimal auction mechanism to decide the cache space allocation and user payments in edge computing.

However, these works focus on content caching policies which are not directly applicable to function caching.
In function caching, one function cannot serve multiple requests simultaneously while content caching can.
Furthermore, our work considers the request distribution and connectivity between edge nodes which renders existing works for content caching inapplicable.
\section{System Model}\label{section:system}

In this section, we present the system model for the serverless request distribution and caching problem.
After that, we present the system design of the proposed framework.

\subsection{Overview of the cross-edge system}
In this paper, we consider an edge service provider deploying serverless computing services on a set of geographically distributed edge nodes, denoted by $\mathcal{V} = \{1,2,...,V\}$. 
Each edge node $v$ is equipped with a certain amount of hardware resources (e.g., memory) illustrated as $U_t^v$.
Similarly, we use $\mathcal{E} = \{1,2,..., E\}$ to represent the set of links between edge nodes.
Also, the set of user requests is denoted by $\mathcal{R} = \{1,2,...,R\}$.
The system works in a time-interval manner spanning across a large period of time $\mathcal{T}$ and each time slot is denoted by $t \in \mathcal{T}$.
Each time interval denotes a decision interval, which is much longer than the processing time of a typical serverless application \cite{Roy2022}.

The service requests are generated by end-users.
A central scheduler orchestrates the service requests to appropriate edge clouds that serve the requests.

\subsection{System design}

Figure~\ref{fig:system} shows the architecture of pCache which is built over Knative, Kubernetes \cite{Kubernet87:online} and Kourier \cite{knativee33:online} tools.
The users first send severless requests to an ingress gateway which is extended from Kourier gateway. 
After that, the serverless requests are sent to pCache's scheduler.
The scheduler makes decisions based on the Algorithm~\ref{alg:distribution} and \ref{alg:procache}.
The cache information is also stored in the scheduler and hence the scheduler jointly factors in the cached containers and the incoming requests.
The decisions are then sent to the Kubernetes' API server through HTTP requests.
The API server will create new containers if required.
In the meantime, the placement decisions are also sent to the ingress gateway and the load balancer will distribute the traffic to an assigned function pod.
Eventually, the function pods process the incoming traffic and return results back to users.

It is worth mentioning that the scheduler keeps a active list and a cache.
The active list contains information of active containers that are serving requests.
The cache contains information of containers that are alive but not serving any requests.
When a container finishes serving a request, the scheduler will decide whether terminating the container based on the probabilistic algorithm~\ref{alg:procache}.

\section{Problem Description}\label{section:problem}
In this work, the system cost of the online serverless function orchestration includes two components: the service latency cost and the container running cost.
All symbols and variables are listed in Table~\ref{tab:var}.

\begin{table}[!tbp]
	\centering
	\normalsize
	\setlength\belowcaptionskip{0ex}
	\caption{Symbols and Variables}
	\label{tab:var}
	\renewcommand\arraystretch{1}            
	\begin{tabular*}{250pt}{ll}
		\toprule	
		\textbf{Symbols and Variables} & \textbf{Description}\\
		\midrule
		$\mathcal{G} = (\mathcal{V}, \mathcal{E})$ & Physical network graph\\
            $\mathcal{V}$ & Set of edge nodes\\
            $\mathcal{E}$ & Set of links\\
            $\mathcal{N}$ & Set of container types\\
            $\mathcal{T}$ & Set of time intervals\\
            $u_n$ & The required hardware resource \\
            & of type $n$ container\\
		$U^v_t$ & The hardware capacity of node $v$\\
		$p^v_n$ & The cost of creating a type $n$ \\
            & container at node $v$\\
		$q^v_n$ & The running cost of type $n$ \\
            & container at node $v$\\
		$d_{v,v'}$  & Communication cost between  \\
            & node $v$ and $v'$\\

		
		$m^{v}_{n,t}$ & Number of type $n$ \\
            & container assigned to node $v$\\
		$m^{v->v'}_{n,t}$ &  Number of type $n$ \\ 
            &container generated at node $v$\\
        & and assigned to node $v'$\\
         $a_n^{v,t}$ & Number of type $n$ containers \\
        & active at node $v$ in time interval $t$\\
         $\lambda_{n,t}^v$ & The total number of requests \\
         & generated at node $v$. \\
         $y_{n,t-1}^v$ & The number of containers to be \\
         & destroyed on node $v$ at end \\
         & of $t-1$. \\
         $k_{v,t}^n$ & The number of requests \\ 
         & remaining to be served\\
         $\alpha$ & Parameter to tune the tradeoff \\
         & between service latency and\\
         & container running cost\\
         $\beta$ & Parameter to tune Zipf distribution \\ 
		\bottomrule
	\end{tabular*}
\end{table}

\subsection{Service latency cost}
The service latency cost consists of two components: the switching cost and the communication cost which are all proportional to the actual latency.

Launching a new serverless container requires transferring the container image containing the serverless application to the hosting edge node and creating a new executive environment which may take up to a few seconds.
We use $p_v^n$ to denote the cost of creating a type $n$ container at edge cloud $v$.
Then, the total container switching cost at time interval $t$ is given by:

\begin{equation}
\label{eq:switchcost}
\begin{aligned}
 \sum_{v \in \mathcal{V}}\sum_{n \in \mathcal{N}} p^v_{n} max \{m^v_{n,t} - a^v_{n,t}, 0\}
\end{aligned}
\end{equation}

where $a^v_{n,t} = max\{a^v_{n,t-1}, m^v_{n,t-1} \} -y^v_{n,t-1}$. 
Let $a^v_{n,t}$ represent the number of active containers of type $n$ on node $v$ at the beginning of time interval $t$. 
$m^v_{n,t}$ denotes the number of required containers of type $n$ on node $v$ at the beginning of time interval $t$. 
$y^v_{n,t-1}$ represents the number of containers to be destroyed on node $v$ in the end of time interval $t-1$.

We use $d_{v,v'}$ to denote the communication cost between two edge nodes $v$ and $v'$ which is proportional to the network communication latency.
The total network communication cost in time slot $t$ is given by:

\begin{equation}
\label{eq:routingcost}
\begin{aligned}
\sum_{v,v' \in \mathcal{V}}\sum_{n \in \mathcal{N}} d_{v,v'}m^{v -> v'}_{n,t}
\end{aligned}
\end{equation}

where $m^{v -> v'}_{n,t}$ represents the number of containers that are generated at edge node $v$ and distributed to edge node $v'$.

Hence, the total service latency cost is given as follows.
\begin{equation}
\label{eq:routingpwr}
\begin{aligned}
C_L(t) = \sum_{v \in \mathcal{V}}\sum_{n \in \mathcal{N}} p^v_{n} max \{m^v_{n,t} - a^v_{n,t}, 0\} +  \\
\sum_{v,v' \in \mathcal{V}}\sum_{n \in \mathcal{N}} d_{v,v'}m^{v -> v'}_{n,t}
\end{aligned}
\end{equation}

\subsection{Container running cost}
The container running cost denotes the hardware resource price paid for running a container.
Hence, the total container running cost in time interval $t$ is given as follows.
\begin{equation}
\label{eq:runpwr}
\begin{aligned}
C_R(t) = \sum_{v \in \mathcal{V}}\sum_{n \in \mathcal{N}} q^v_n m^v_{n,t}
\end{aligned}
\end{equation}

where $q^v_n(t)$ represents the system cost of running a container $n$ in edge cloud $v$ in one time slot, attributed to hardware resources paid for running a container.

Hence, the total cost of the system is a sum of the service latency cost and the container running cost.

\begin{equation}
\label{eq:allpower}
\begin{aligned}
  \  C_L(t) + \alpha C_R(t) 
\end{aligned}
\end{equation}

where the parameter $\alpha$ is used to tune the tradeoff between service latency cost and container running cost. 
Note that $\alpha q^v_n \leq p^v_{n}$ should always be followed, otherwise caching containers will be more expensive than creating new containers, making the function caching unhelpful.

\subsection{The system cost minimization problem}
In this work, we aim to devise a cost-efficient request distribution and caching framework for serverless edge computing.
To this end, we formulate a joint optimization on container placement and container caching, aiming at minimizing the overall system cost.

\begin{problem}
\begin{equation}
\label{eq:allpower}
\begin{aligned}
min \quad \sum_{t \in \mathcal{T}} (C_L(t) + \alpha C_R(t))
\end{aligned}
\end{equation}
\end{problem}

s.t.

\begin{equation}
\label{con:capacity}
\begin{aligned}
 \sum_{n \in \mathcal{N}}  u_{n}  m^v_{n,t} \leq U^v_t,
 \forall v \in \mathcal{V},\forall t \in \mathcal{T}
\end{aligned}
\end{equation}

\begin{equation}
\label{con:one}
\begin{aligned}
 \sum_{n \in \mathcal{N}}  m^{v -> v'}_{n,t} = \lambda_{n,t}^{v},
 \forall v,v' \in \mathcal{V},\forall t \in \mathcal{T}
\end{aligned}
\end{equation}

\begin{equation}
\label{con:y}
\begin{aligned}
y^v_n(t)\in [0, max\{a_v^{n,t}, m_v^{n,t} \}] \quad \forall t \in \mathcal{T}, \forall n \in \mathcal{N}, v \in \mathcal{V}
\end{aligned}
\end{equation}

\begin{equation}
\label{con:range}
\begin{aligned}
 m^v_{n,t}\in \{0,1,...,\lambda^v_n\}, \quad \forall t \in \mathcal{T}, \forall n \in \mathcal{N}, v \in \mathcal{V}
\end{aligned}
\end{equation}

Eq.~\ref{con:capacity} guarantees that the allocated hardware resource does not exceed the maximum capacity on each node.
Eq.~\ref{con:one} ensures that every request is distributed to only one node.
Eq.~\ref{con:y} ensures that the number of destroyed containers is less than the number of active containers.
Eq.~\ref{con:range} represents the integrality constraint for the number of distributed containers $m^v_{n,t}$.

\subsection{NP hardness}

We show that the Generalized Assignment Problem (GAP) \cite{Chu1997}, which is proven to be NP-hard, can be reduced to a simplified version of the proposed problem.
The GAP problem is assigning a number of $K$ jobs to a set of $J$ agents, aiming to minimize the total cost.
Let $\omega_j^k$ denote the size of job $k$ for agent $j$ to perform the job.
Let $d_j^k$ represent the cost of running job $k$ for agent $j$.
Also, we use $\Omega_j^k$ to denote the capacity of agent $j$ and binary variable $x_j^k$ to present whether job $k$ is assigned to agent $j$.
Finally, the GAP problem can be formulated as follows.

\begin{problem}
\begin{equation}
\label{eq:problem2}
\begin{aligned}
 min \sum_{k \in \mathcal{K}}\sum_{j \in \mathcal{J}} d_j^k x_j^k
\end{aligned}
\end{equation}
\end{problem}

s.t.

\begin{equation}
\begin{aligned}
 \sum_{j \in \mathcal{K}}  x_j^k = 1, \forall k
\end{aligned}
\end{equation}

\begin{equation}
\begin{aligned}
 \sum_{k \in \mathcal{K}}  \omega_j^k x_j^k = \Omega_j^k, \forall j
\end{aligned}
\end{equation}

\begin{equation}
\begin{aligned}
x_j^k \in [0,1], \forall j, \forall k
\end{aligned}
\end{equation}

Now we show the equivalence of the GAP problem and the simplified version of our proposed problem.
If a type $n$ request is generated once on edge node $v$ ($\sum_{v \in \mathcal{V}} m^v_{n,t} =1$), the communication cost is 0 because we can distribute the request to edge node $v$, namely where the request origins.
Also, we terminate a container immediately after serving a request which means $a_{n,t}^v = 0$.
Then, the proposed problem is converted to optimize the switching and running costs.
We formulate the simplified problem of cost minimization as follows.

\begin{problem}
\begin{equation}
\label{eq:problem3}
\begin{aligned}
 \sum_{v \in \mathcal{V}}\sum_{n \in \mathcal{N}} (p^v_{n} + \alpha q^v_n ) m^v_{n,t}
\end{aligned}
\end{equation}
\end{problem}

s.t.

\begin{equation}
\begin{aligned}
 \sum_{v \in \mathcal{V}}  m_{n,t}^v = 1, \forall k, \forall t
\end{aligned}
\end{equation}

\begin{equation}
\begin{aligned}
 \sum_{n \in \mathcal{N}}  u_n m_{n,t}^v \leq U_t^v, \forall v, \forall t
\end{aligned}
\end{equation}

\begin{equation}
\begin{aligned}
m_{n,t}^v \in [0,1], \forall v, \forall n, \forall t
\end{aligned}
\end{equation}

If we map job $k$, agent $j$, job assignment cost $d_j^k$, the job size $\omega_j^k$ and the resource capacity $\Omega_j^k$ in GAP problem to function $n$, edge node $v$, system cost $p^v_{n} + \alpha q^v_n$, function resource demand $u_n$ and resource capacity $U_t^v$, the equivalence of \textbf{Problem} 2 and \textbf{Problem} 3 is achieved.
Therefore, the GAP problem is a special case of the proposed problem.

As proved above, the proposed request distribution problem is NP-hard.
Thus, we propose Algorithm~\ref{alg:distribution} in the next section and prove the theoretical gap of performance.

\section{Online Optimization for System Cost}\label{section:solution}

In this section, we first present a request distribution algorithm that assigns serverless requests to suitable edge nodes.
The particular reason for this circumstance is that we jointly consider the remaining resource capacity and cached containers on each edge node.
Further, we introduce a probabilistic caching algorithm to decide what container to terminate when the resource capacity is close to the limit.

\subsection{Request distribution}
\begin{algorithm}[tb]
	\caption{Request Distribution}
	\label{alg:distribution}
	\normalsize
	\SetKwInOut{Input}{Input}
	\SetKwInOut{Output}{Output}
	
	\ForEach{$t \in \mathcal{T}$}
	{
		\ForEach{$v \in \mathcal{V}$}
		{
			\ForEach{$n \in \mathcal{N}$}{
				
				Sort all nodes $v'$ by $d_{v,v'}$ in ascending order in the sorted list $\mathcal{V'}$;\\
				Update the probability of type $n$ function acc. to Equation~\ref{eq:prob}.\\
					
				\uIf{$\lambda_{v,t}^n \leq a_{v,t}^n$}
				{
					Assign all $\lambda_{v,t}^n$ requests to current node $v$.\\
					$m_{v,t}^n \leftarrow \lambda_{v,t}^n$.\\
				  $k_{v,t}^n \leftarrow 0$ .\\
				}
				\ElseIf{$\lambda_{v,t}^n \geq a_{v,t}^n$}
				{
		 			$m_{v,t}^n \leftarrow a_{v,t}^n$.\\
		 			$k_{v,t}^n \leftarrow \lambda_{v,t}^n -  a_{v,t}^n$.\\
		 			
		 			\ForEach{$v' \in \mathcal{V'}$}{
				  
						\If{$d_{v,v'}$ $\leq$ $p_n^v$} 
						{
							\uIf{$a_{v',t}^n \geq k_{v,t}^n$}
							{
								$m_{v',t}^n \leftarrow k_{v,t}^n$.\\
							    $k_{v,t}^n \leftarrow 0$. \\
								break;
							}
							\Else
							{   
								$m_{v',t}^n \leftarrow a_{v',t}^n$.\\
								$k_{v,t}^n \leftarrow k_{v,t}^n -a_{v',t}^n $. \\
							}
						}
			
				    }
			    
			        \If{$k_{v,t}^n \neq 0 $} 
			        {
			        	Call Algorithm~\ref{alg:procache}, instantiate $k_{v,t}^n $ new containers at current node $v$.\\
			        }
		    
				}
		    }
	   }
   }
\end{algorithm}

Intuitively, a request generated at a particular edge node should be served by cached containers at the current node if the current edge node suffices hardware resources.
In this case, the switching cost and communication cost are reduced to 0.
Otherwise, a new container needs to be created at the current edge node at the cost of the switching cost or the request can be distributed to a neighbor edge node at the cost of extra communication cost.

As illustrated in Algorithm~\ref{alg:distribution}, when a request arrives at an edge node $v$,
all edge nodes are sorted in ascending order based on the communication cost $d_{v,v'}$ to the current node.
If the number of type $n$ requests $\lambda_{v,t}^n$ is smaller than the number of cached containers $a_{v,t}^n$ at the current node $v$, all the requests are distributed to the current edge node and served by cached containers.
In other words, if the current edge node has enough cached containers, we assign all requests to the current node $v$.
Otherwise, we distribute the unprocessed requests $k_{v,t}^n$ to neighbor nodes $v' \in \mathcal{V}$ that have sufficient cached containers if the communication cost $d_{v,v'}$ is less than the switching cost $p_n^v$.
The particular reason for this circumstance is that offloading requests to neighbor nodes is more beneficial than creating new containers at the current node in this case.
If there still are unprocessed requests after this, we create new containers in the current edge node at the cost of extra switching cost by using Algorithm~\ref{alg:procache}.
Since the edge node is constrained by resource capacity, cached containers need to be destroyed for new containers when the resource capacity is insufficient in edge nodes.
The cached containers are destroyed based on a probabilistic algorithm as shown in Algorithm~\ref{alg:procache}.

\begin{algorithm}[tb]
  \caption{Probabilistic Caching}
  \label{alg:procache}
  \normalsize
  \SetKwInOut{Input}{Input}
  \SetKwInOut{Output}{Output}

    \While{$\sum_{n \in \mathcal{N}}  u_n m_{v,t}^n \textgreater U_{t}^v$ }
    {
      Calculate the probability $P_n(t)$ for each type of container $n$;\\
      Random a container type based on $P_n(t)$;\\
      Destroy a container based on the random result;\\
    }
    Create a new type $n$ container for the request;\\
\end{algorithm}

\begin{theorem}
The worst-case system cost for handling a type $n$ request is bounded by $\max\{1 + \frac{p_n^v}{\alpha q_n^v},  1 + \frac{d_{v,v'}}{\alpha q_n^v}  \}$.
\end{theorem}

\begin{proof}
The offline optimal solution of distributing a type $n$ request is assigning the request to a cached container in the current edge node.
In this case, the switching cost is 0 as a cached container is used.
Similarly, the communication cost is also 0 since the request is served at the edge node where it is generated.
Thus, the system cost of a type $n$ request is lower bounded by:
\begin{equation}
\centering
OPT(n) \geq \alpha q_n^v
\end{equation}

After that, we consider the worst case produced by Algorithm~\ref{alg:distribution}.
The worst case includes two cases. The former is that we create a new container at the cost of switching cost $p_n^v$.
The latter is that we offload the request to a nearby edge node at the cost of communication cost $d_{v,v'}$.
Hence, we define the maximum system cost of our algorithm as:
\begin{flalign}
\centering
WORST(n) = 
\begin{cases}
p_n^v + \alpha q_n^v& if \quad d_{v,v'} \geq p_n^v\\
d_{v,v'} + \alpha q_n^v & if \quad d_{v,v'} <  p_n^v\\
\end{cases}
\end{flalign}

We define $f(n)$ as $f(n) = WORST(n)/OPT(n)$.
Hence, the maximum of $f(n)$ denotes the worst-case competitive ratio.
We obtain $f(n)$ for the system cost to serve a request as:

\begin{flalign}
\centering
f(n) = 
\begin{cases}
1 + \frac{p_n^v}{\alpha q_n^v}& if \quad d_{v,v'} \geq p_n^v\\
\\
1 + \frac{d_{v,v'}}{\alpha q_n^v}& if \quad d_{v,v'} <  p_n^v
\end{cases}
\end{flalign}

Hence, the maximum $f(n)$ achieved by the proposed algorithm is given by
$\max\{1 + \frac{p_n^v}{\alpha q_n^v},  1 + \frac{d_{v,v'}}{\alpha q_n^v}  \}$.
\end{proof}



    

\subsection{Probabilistic function caching}
In this section, we elaborate on the proposed probabilistic caching policy inspired by a web caching method \cite{Cherkasova1998}.
The key insight of this paper is that different types of containers should have different probabilities of being evicted from the cache.
This is because the resource footprint and invocation frequency vary between different types of containers.
When an edge node does not have sufficient hardware resources to create new containers, we use the probabilistic caching policy to decide which container in the cache to destroy.
Another benefit of probabilistic caching is to not be overly aggressive with caching the most popular container.
By offering a probability to cache less popular containers, we intend to be fair with different kinds of containers.

As illustrated in Algorithm~\ref{alg:procache}, when the required amount of resources exceeds the remaining resource capacity, the proposed algorithm will calculate probabilities $P_n(t)$ for each type of function.
Then, we randomly select a container type based on the set of $P_n(t)$.
Thus, we destroy a container that is selected.
Finally, we repeat this process until there are sufficient resources for creating new containers.

\textbf{Probability Calculation}
We propose a probabilistic caching policy, which is contextual, to determine which container to destroy in the cache when an edge node is running close to the capacity.
The probability of a type $n$ function being evicted from the cache at time slot $t$ is given by:

\begin{equation}
	\label{eq:prob}
	\begin{aligned}
		P_n(t)= \frac{ \frac{u_n}{f_n  + t_n}} {\sum_{n \in N}  \frac{u_n}{f_n + t_n}}
	\end{aligned}
\end{equation}

\textbf{Size}
Size $u_n$ is the memory footprint of a container as containers are cached in memory.
Since the probability of eviction is proportional to the size, large containers are more likely to be evicted than small containers.

\textbf{Frequency}
Frequency $f_n$ represents the number of times a particular function is invoked until the current time interval. 
The frequency is updated every time a container is created or destroyed.
The probability is inversely proportional to the frequency and hence frequently invoked functions have low probability to be evicted from the cache.


\textbf{Recency}
We use $t_n$ to denote the last time a type $n$ function is invoked.
$t_n$ reflects the recency a function was invoked.
The function, that is recently used, will have a low probability to be evicted from the cache.

\textbf{Serverless-specific considerations}
Since the images for all containers of a function type $n$ are identical, it is reasonable to assume that those containers have identical container sizes \cite{Fuerst2021}.
Thus, any of the identical containers can be terminated.
\section{Performance Evaluation} \label{section:performance}

We evaluate the performance of pCache over simulation and implementation in Knative \cite{knativee33:online}, which is an open-source serverless platform that powers enterprise-level solutions.
All experiments are repeated 10 times and we report the average of them.

\begin{figure}[tbhp]
    \centering
    \subfloat{\includegraphics[width=0.4\textwidth]{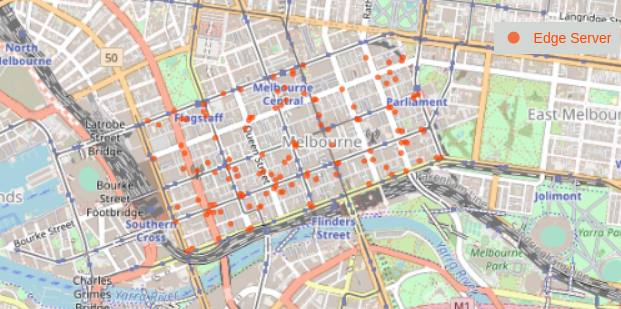}}
    \caption{Edge servers in Melbourne CBD area}
    \label{fig: topo}
\end{figure}

\begin{figure*}[htbp]
    \centering
    \subfloat[$\beta=0.5$]{\includegraphics[width=0.3\textwidth]{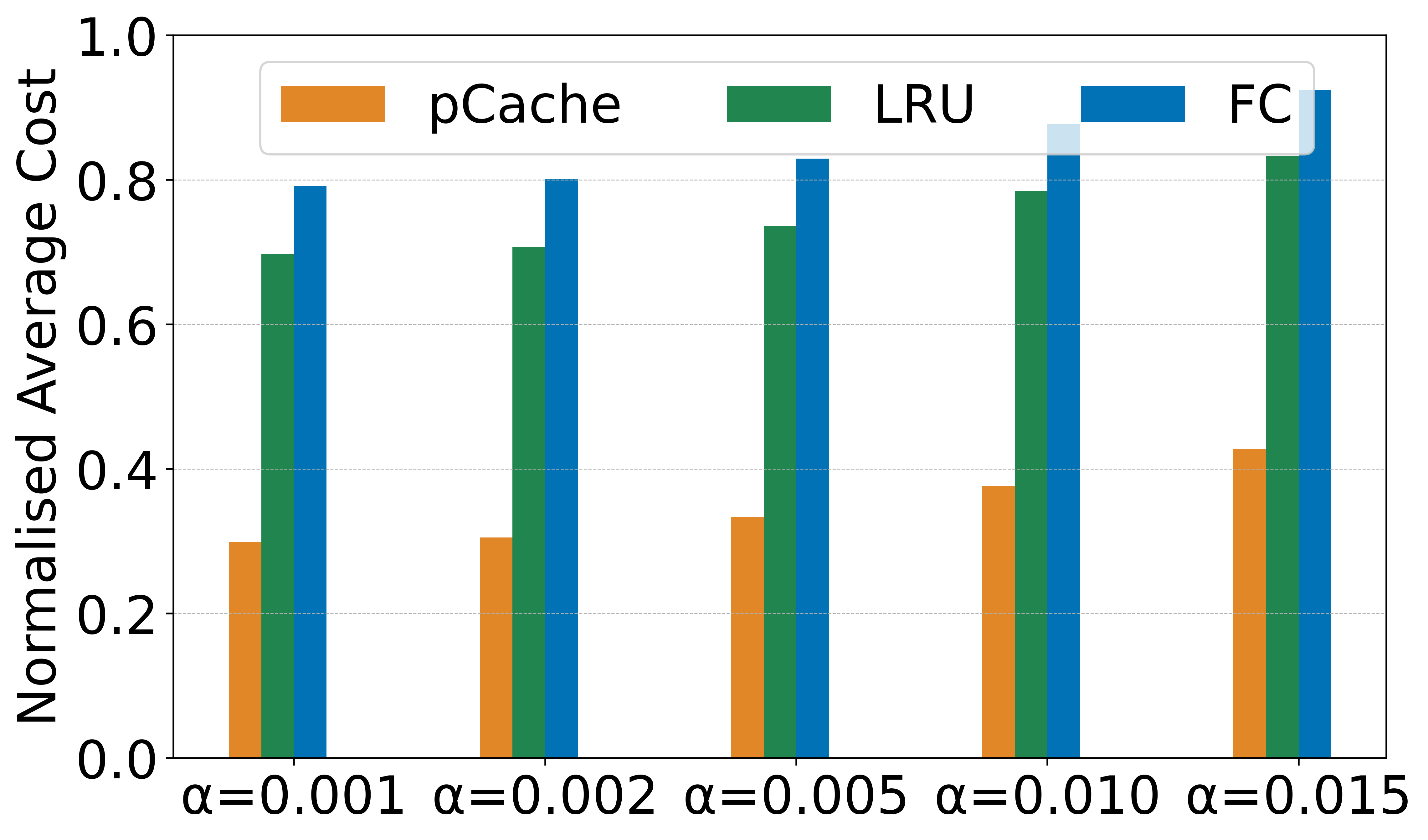}}
    \hfill
    \subfloat[$\beta=1.0$]{\includegraphics[width=0.3\textwidth]{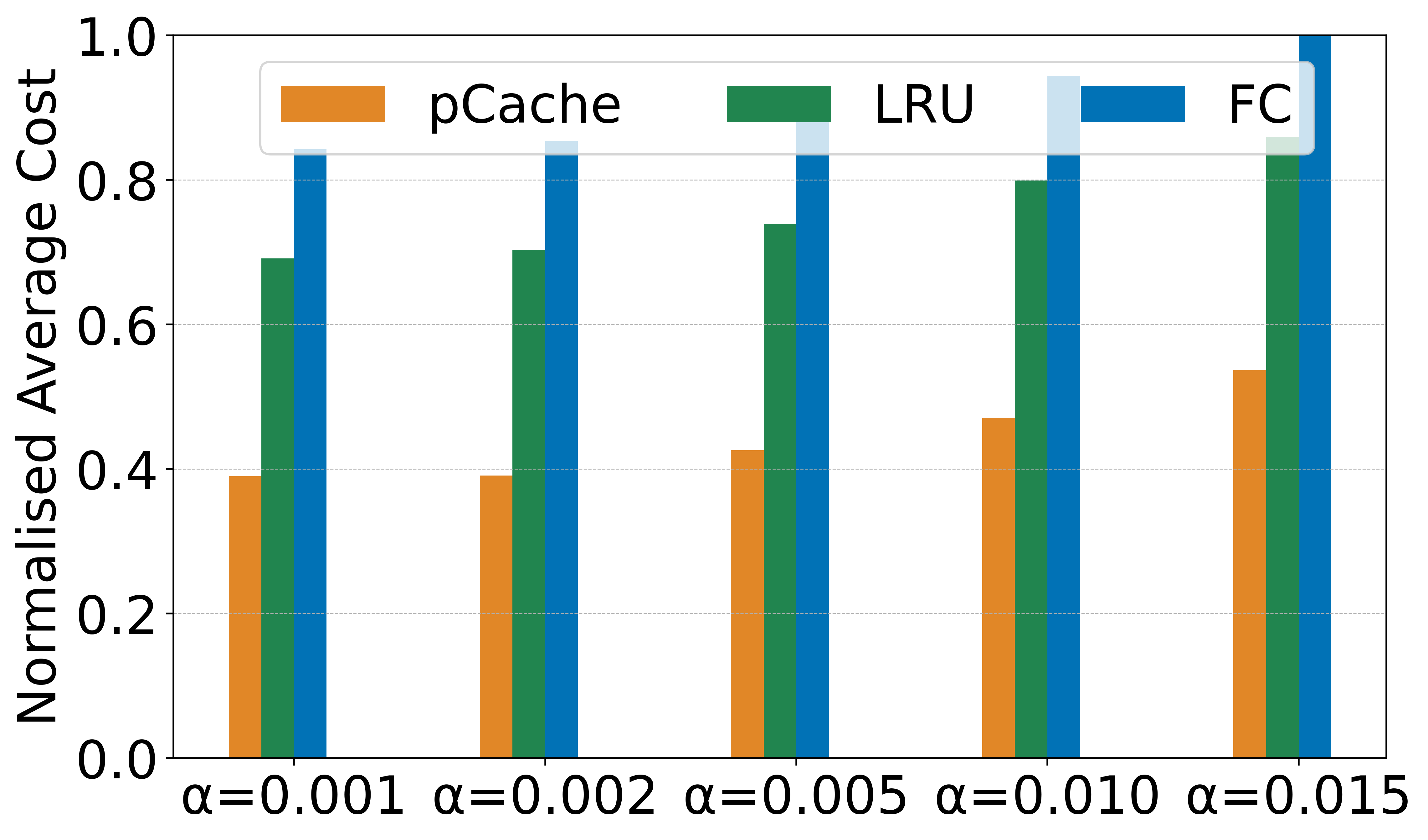}}
    \hfill
    \subfloat[$\beta=1.5$]{\includegraphics[width=0.3\textwidth]{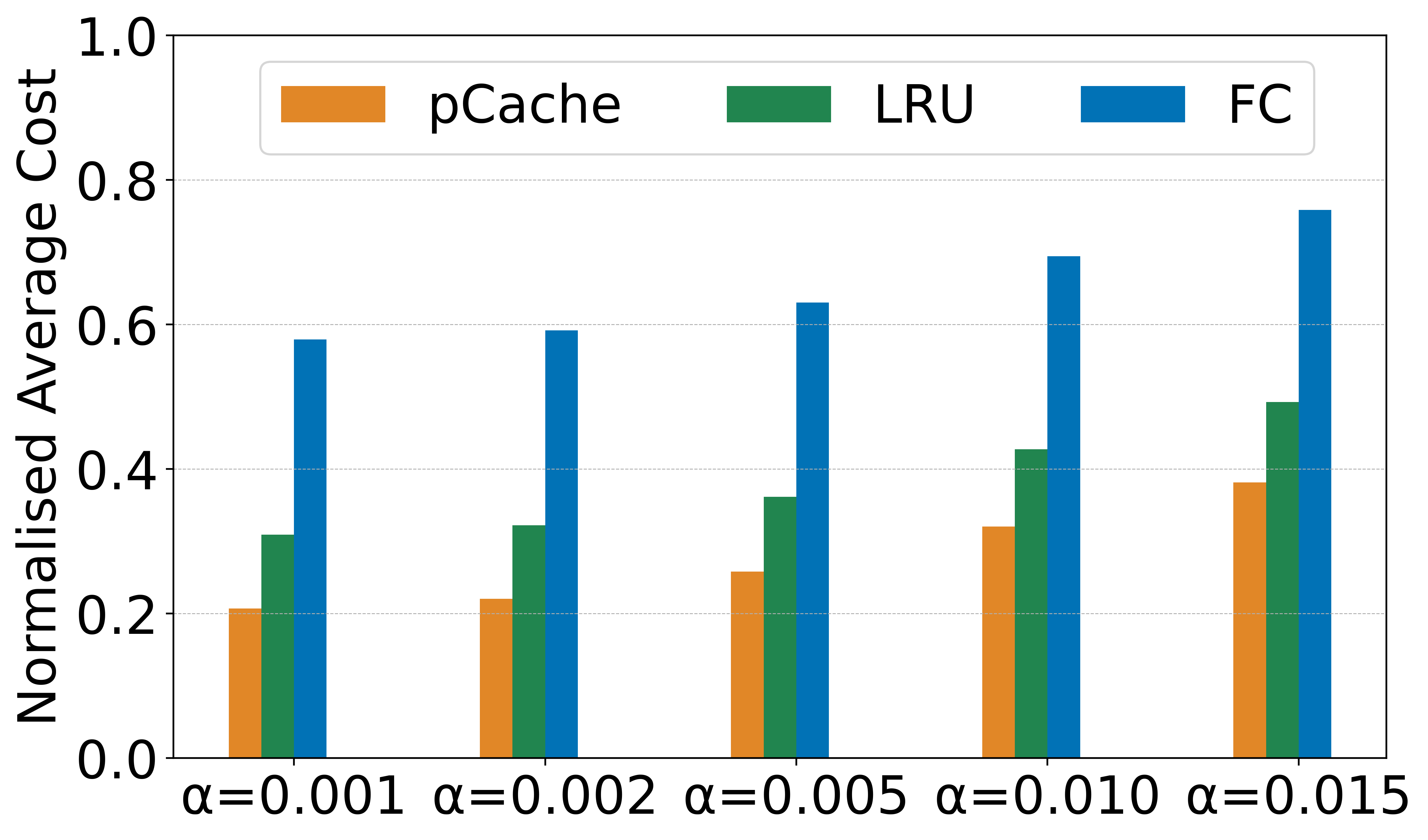}}
    \hfill
   \caption{Average cost in simulation}
    \label{fig: average-cost-ns3}
\end{figure*}

\begin{figure*}[htbp]
  \centering
    \subfloat[$\beta=0.5$\label{fig: boxplot-ns-0.5}]{\includegraphics[width=0.3\textwidth]{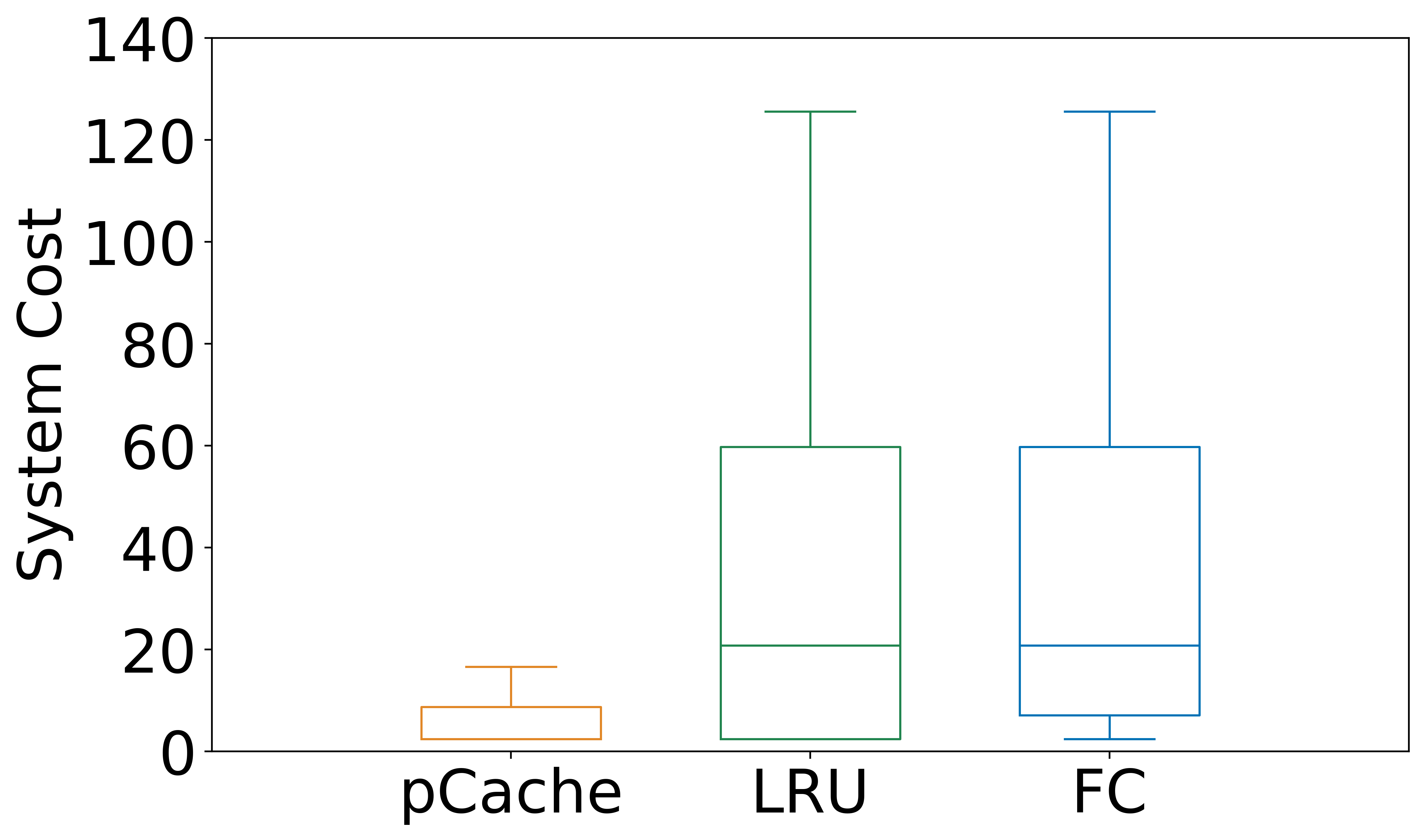}}
    \hfill
    \subfloat[$\beta=1.0$\label{fig: boxplot-ns-1}]{\includegraphics[width=0.3\textwidth]{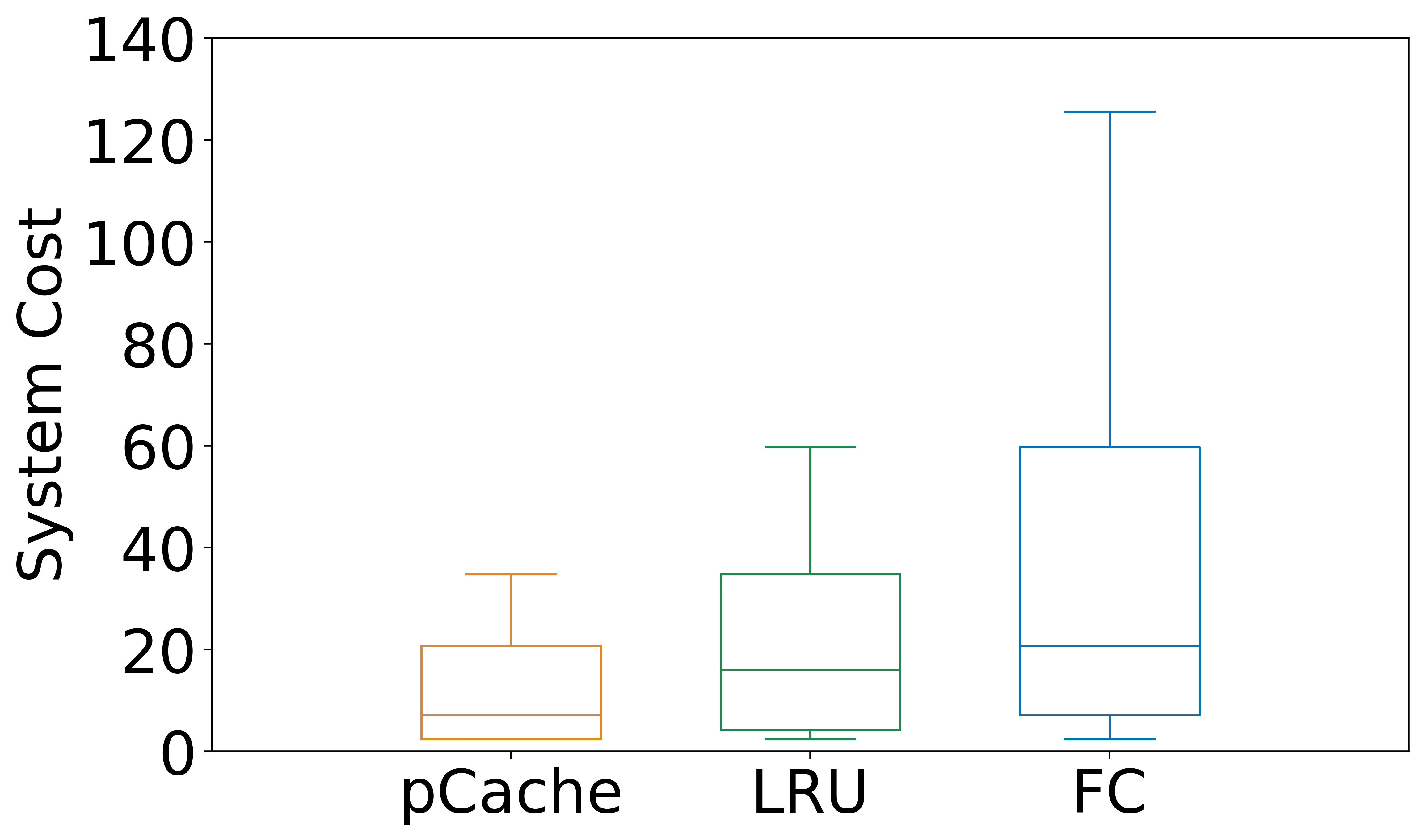}}
    \hfill
    \subfloat[$\beta=1.5$\label{fig: boxplot-ns-1.5}]{\includegraphics[width=0.3\textwidth]{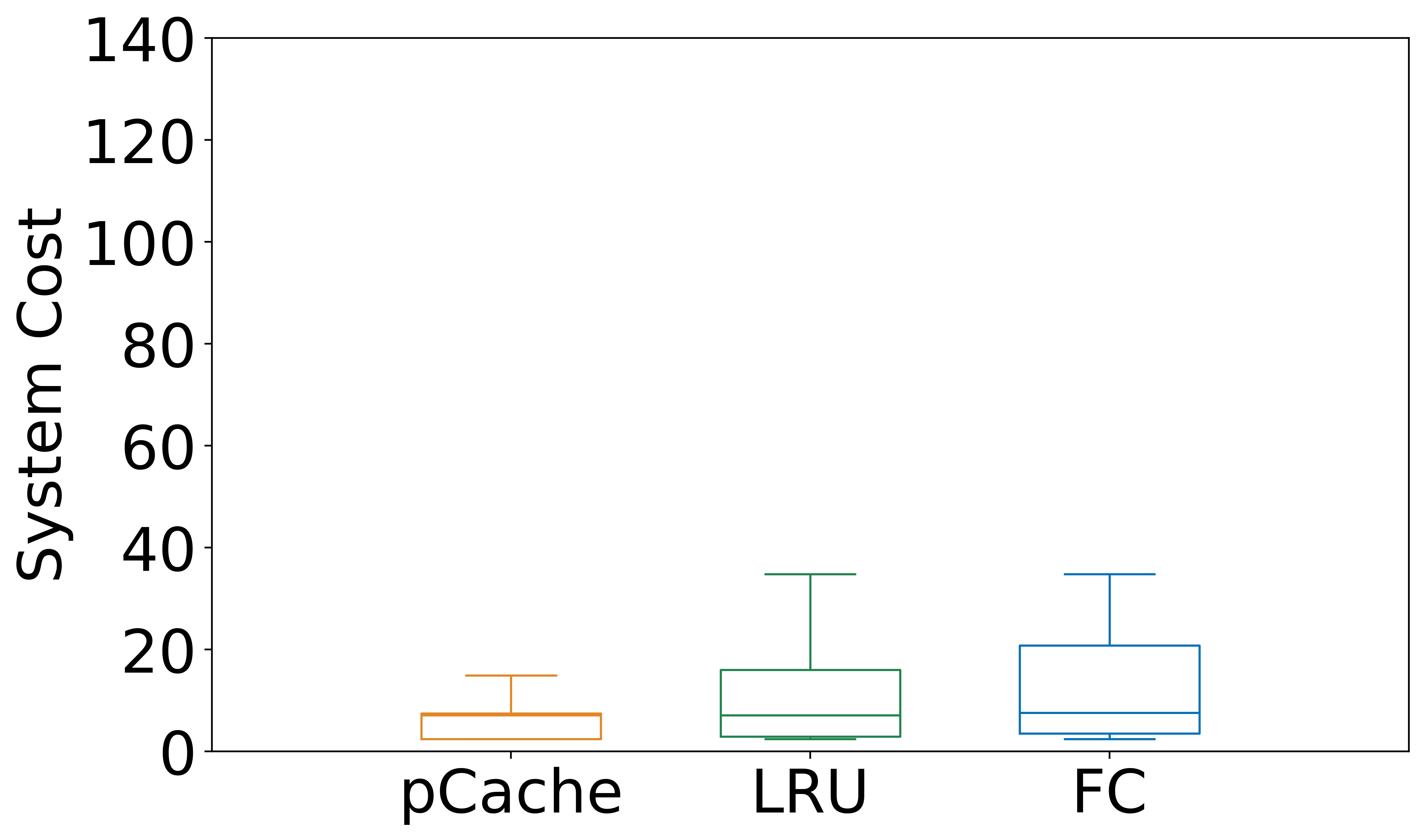}}
    \hfill
    \caption{Distributions of system cost in simulation}
    \label{fig: boxplot-ns3}
\end{figure*}

\subsection{Simulation setup}

\begin{figure}[htbp]
    \centering
    \subfloat{\includegraphics[width=0.36\textwidth]{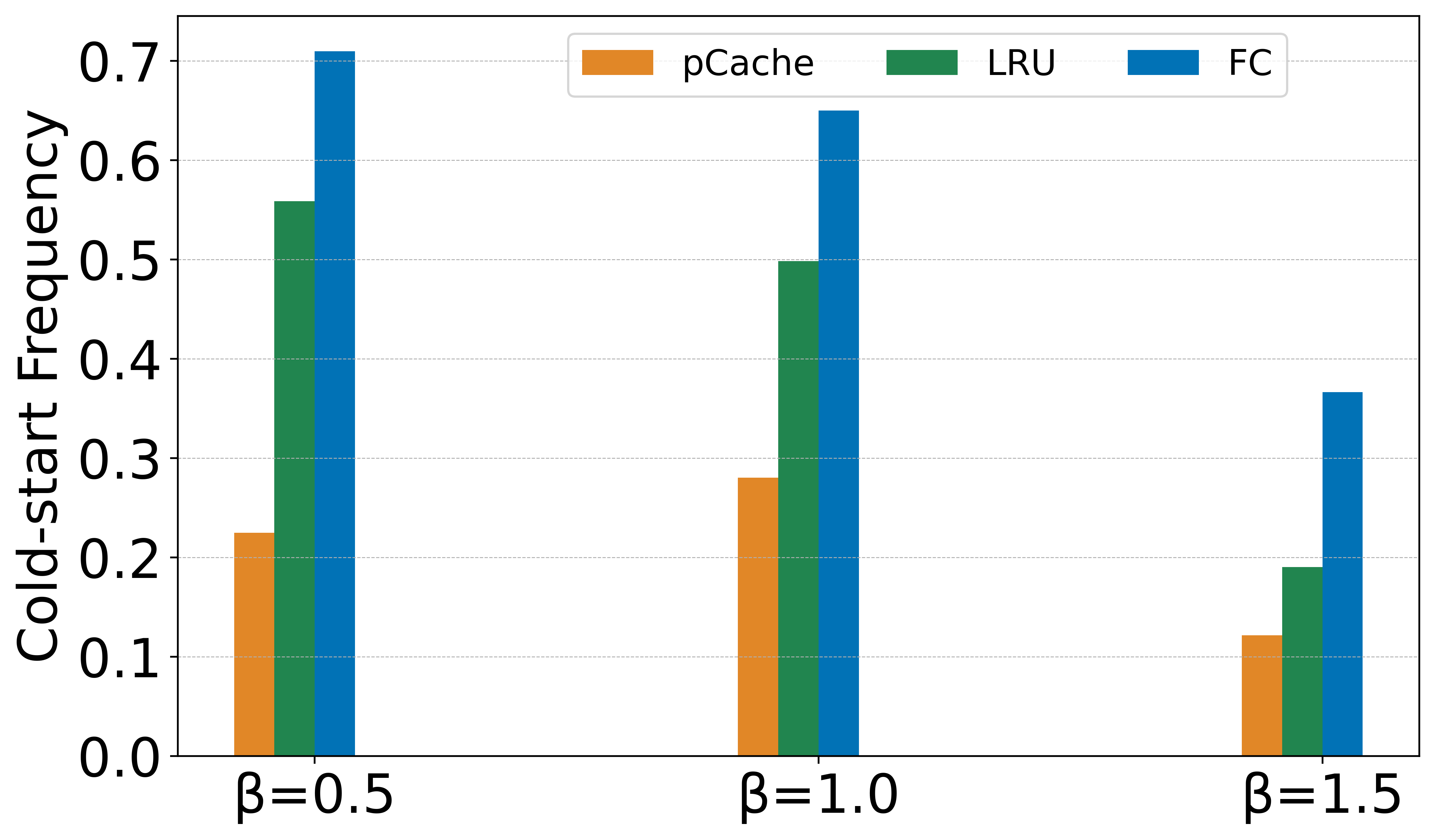}}
    \hfill
    
    \caption{Cold-start frequency in simulation}
    \label{fig: cold-ns3}
\end{figure}

To evaluate the performance of the proposed algorithms, we first conducted extensive simulations on a server with $105$ GB RAM and an Intel(R) Xeon(R) E5645 processor with $24$ cores.


\textbf{Topology:} We use the EUA dataset \cite{Lai2018} which includes the
information on edge servers in the Melbourne CBD area.  
The topology consists of 125 edge nodes as illustrated in Figure~\ref{fig: topo}.

\textbf{Requests:} We use the Azure dataset \cite{azure2019} to generate the serverless requests. This dataset contains the invocations of functions on
Microsoft Azure for 14 days. The workload at each node is generated
using the Zipf distribution \cite{Pan2022} with the exponent ranging from 0.5
to 1.5. We use the top four applications in the Azure dataset and map them to four containers shown in Table \ref{tab::container}. 

\begin{table}[tbhp]
  \centering
  \begin{tabular*}{175 pt}{|l|r|} \hline	 
    \textbf{Application Name} & \textbf{Memory Size}\\ \hline
    Web Server  & 55 MB\\ \hline
    File Processing  & 158 MB\\ \hline
    Supermarket Checkout & 332 MB \\ \hline
    Image Recognition  & 92 MB\\ \hline
  \end{tabular*}
  \caption{Function instances}\label{tab::container}
\end{table}

\begin{figure*}[t]
 \subfloat[$\beta=0.5$]{\includegraphics[width=0.30\textwidth]{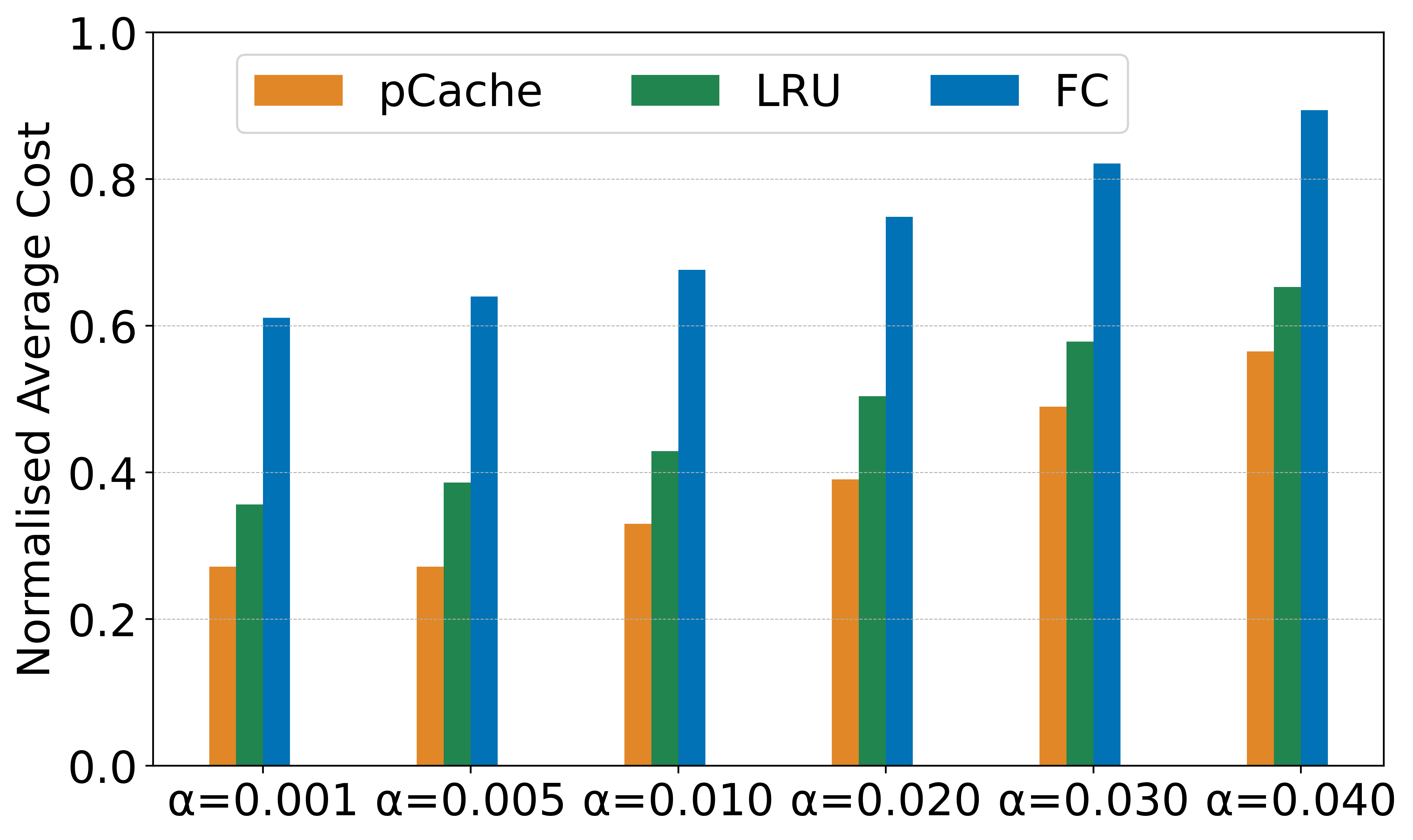}}
    \hfill
    \subfloat[$\beta=1.0$]{\includegraphics[width=0.30\textwidth]{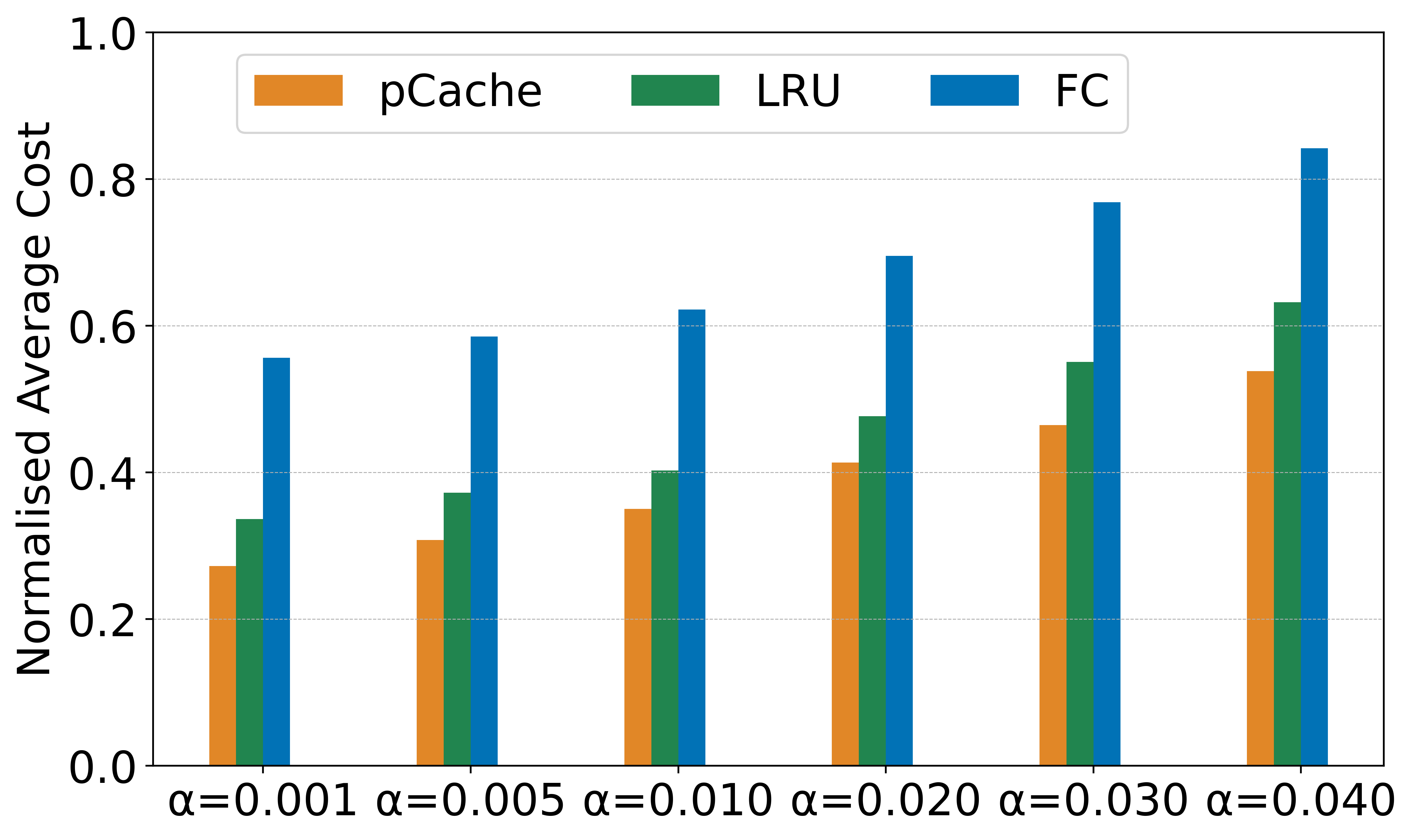}}
    \hfill
    \subfloat[$\beta=1.5$]{\includegraphics[width=0.30\textwidth]{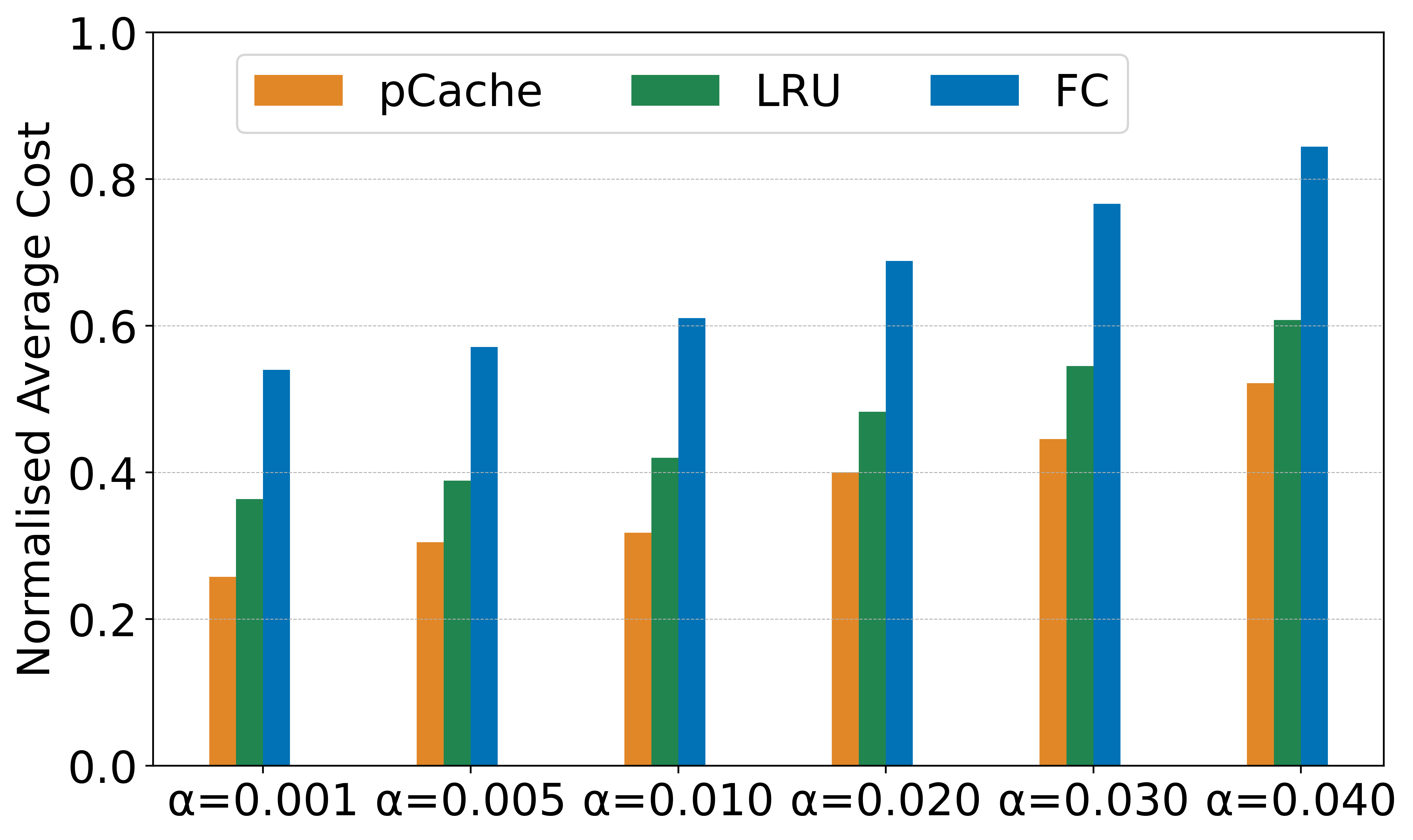}}
    \hfill
    \caption{Average cost in Knative}
    \label{fig: average-cost-k8s}
\end{figure*}

\begin{figure*}[t]
 \subfloat[$\beta=0.5$\label{fig: boxplot-k-0.5}]{\includegraphics[width=0.30\textwidth]{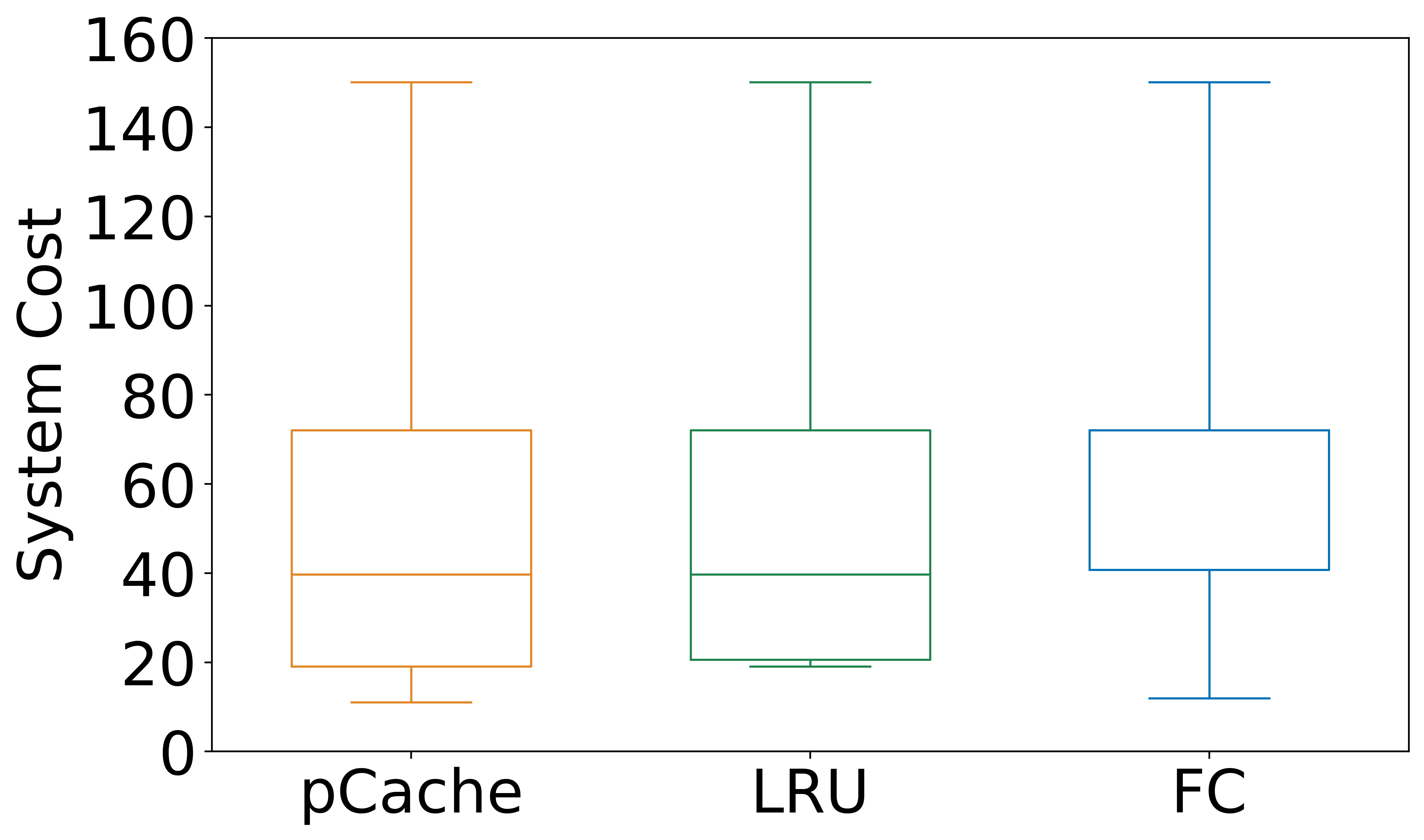}}
    \hfill
    \subfloat[$\beta=1.0$\label{fig: boxplot-k-1}]{\includegraphics[width=0.30\textwidth]{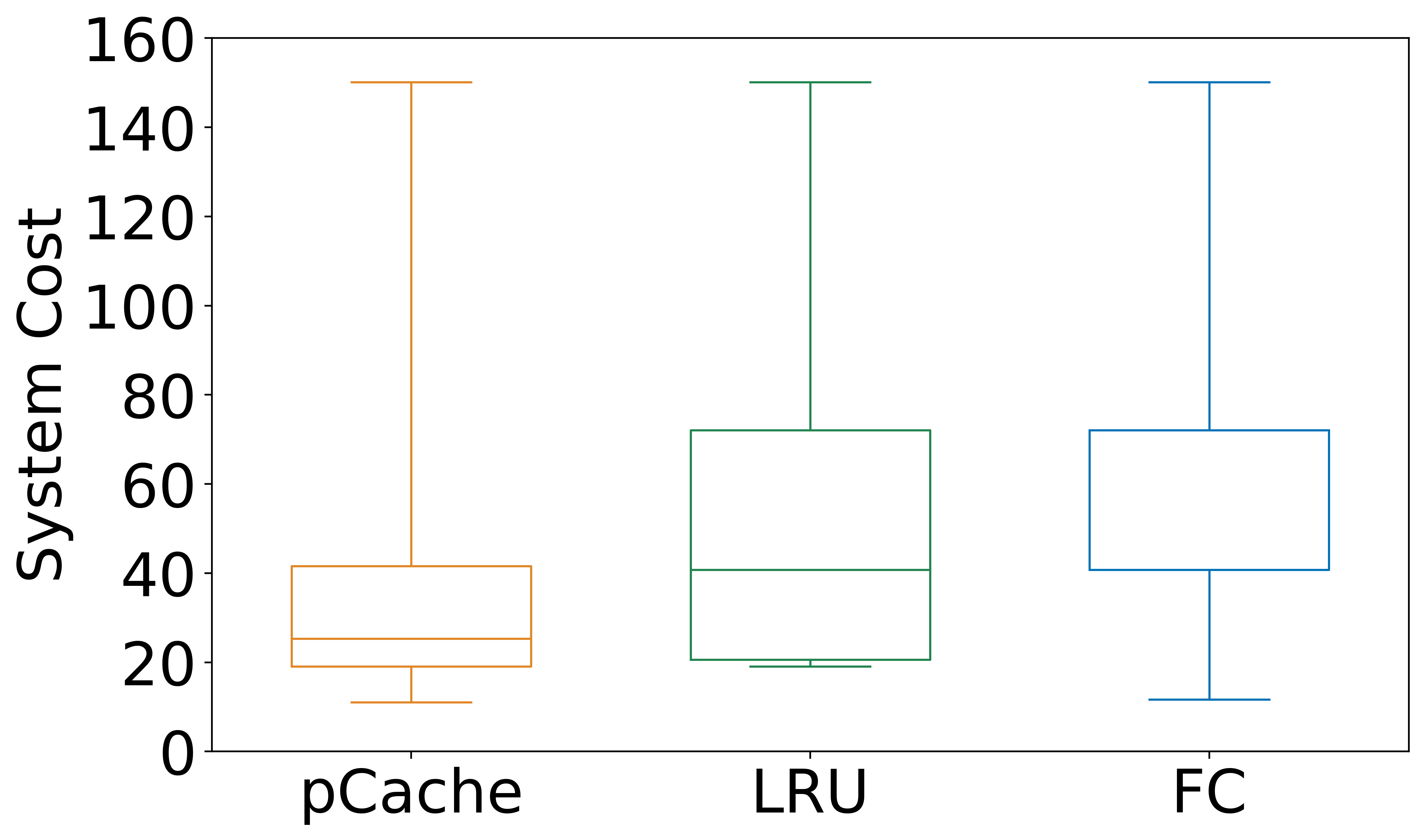}}
    \hfill
    \subfloat[$\beta=1.5$\label{fig: boxplot-k-1.5}]{\includegraphics[width=0.30\textwidth]{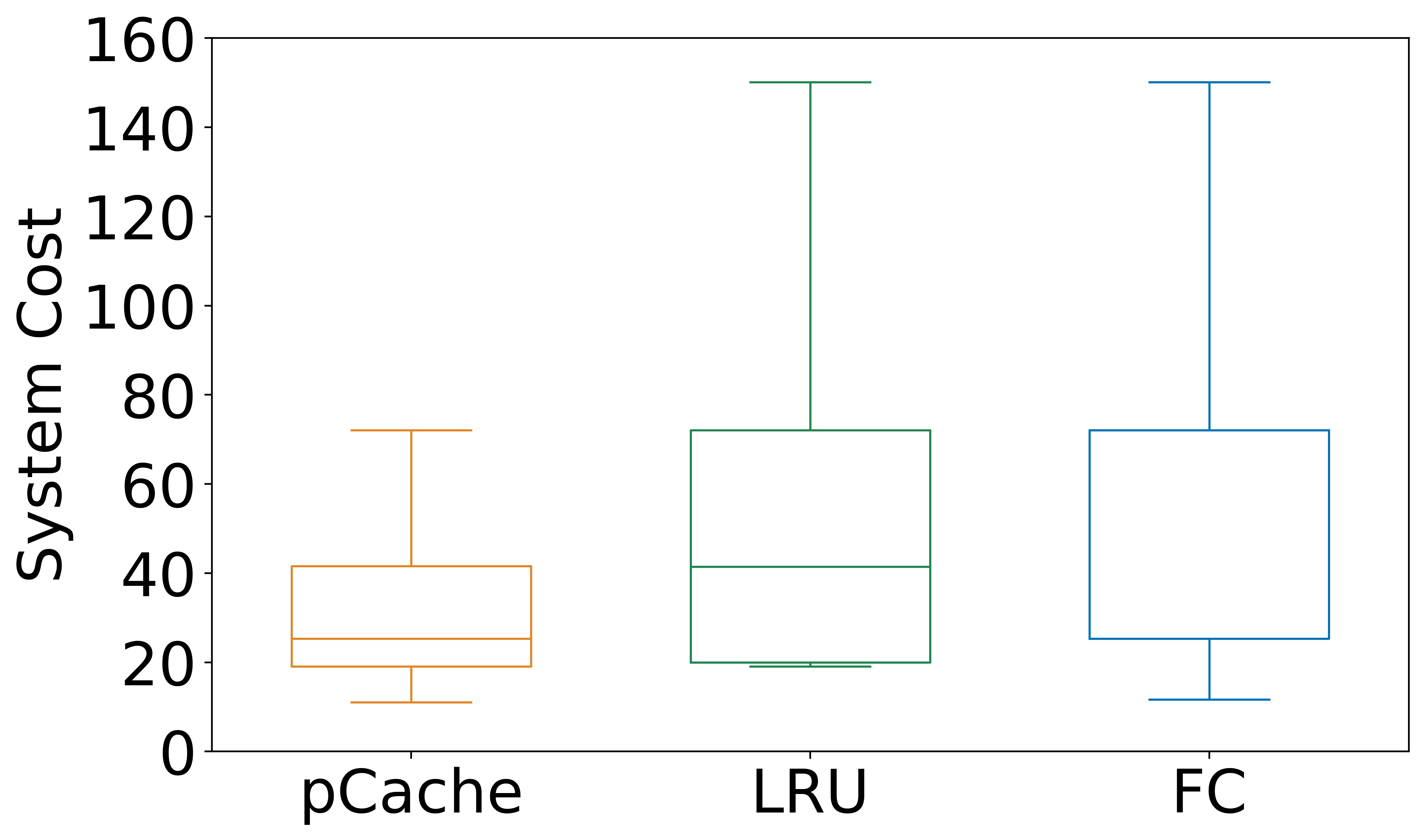}}
    \hfill
    \caption{Distributions of system cost in Knative}
    \label{fig: boxplot-knative}
\end{figure*}

\textbf{Containers:} We have built four types of containers obtained from
AWS Lambda functions, as shown in Table \ref{tab::container}. The memory resource allocated to each container is between 55 and 332 MB.

\textbf{Switching and running cost:}
According to \cite{Manco2017}, we set the container switching cost $p_n^v$ inversely proportional to the CPU frequency of node $v$.
Similarly, we set the container running cost $q_n^v$ proportional to the CPU frequency of edge node $v$. 
Also, $p_n^v$ and $q_n^v$ are proportional to the container size $u_n$.

\textbf{Performance Benchmarks:} 
We select two approaches that are widely used in real-world systems.
Least Recently Used (LRU) caching \cite{Mvondo2021} evicts the container that has not been used for the longest time.
Fixed Caching (FC) is widely used in AWS Lambda \cite{Serverle75:online}. It keeps a
container alive for a fixed period of time. In the experiments, the
caching period was set to 10 minutes, and all the algorithms have
the same cache size.

\subsection{Performance in simulation}

In the simulation, we show the normalized system costs achieved by two benchmarks and our proposed approach in Fig.~\ref{fig: average-cost-ns3} in which (a), (b) and (c) respectively present the results achieved by setting $\beta$ as 0.5, 1.0 and 1.5 when considering $\alpha$ as \{0.001, 0.002, 0.005, 0.010, 0.015\}.

\textbf{Performance summary in simulation:}
pCache reduces the average system cost by up to 57.2$\%$ and up to 62.1$\%$ compared to LRU and FC, respectively.
For cold-start frequency, pCache outperforms LRU and FC by up to 60.8$\%$ and 69.1$\%$, respectively.
The rationale is that the cached containers of pCache are more frequently reused which can be justified by the cold-start frequency.
This actively demonstrates the benefits of capturing context for function caching in serverless computing.

\textbf{Impact of parameter $\alpha$:} 
As aforementioned, $\alpha$ is a parameter to tune the trade-off between service latency cost and container running cost.
We set $\alpha$ from 0.001 to 0.015. When $\alpha$ is greater than 0.015, the container running cost will be larger than the container switching cost, implying that caching functions will always lead to more cost than creating new containers and hence function caching cannot provide any benefits.
From Fig.~\ref{fig: average-cost-ns3}, we observe that the parameter $\alpha$ has a noticeable impact on the overall system cost. 
Overall, when $\alpha$ ranges from 0.001 to 0.015, pCache achieves the best performance where the normalized system cost ranges from 0.3 to 0.43.
In contrast, the system cost of LRU ranges from 0.7 to 0.83.
The rationale is that LRU only considers the invocation frequency when evicting functions from the cache.
FC achieves 0.79 to 0.92 in normalized system cost because FC cannot dynamically adapt to time-varying workloads.

\textbf{Impact of parameter $\beta$:}
As aforementioned, the users at each edge node will randomly generate requests conforming to the Zipf-$\beta$ popularity law \cite{Shukla2018}.
As illustrated in Figure~\ref{fig: average-cost-ns3}, when $\beta$ ranges from 0.5 to 1.5, pCache always achieves the best performance in system cost.
This is because pCache incorporates several characteristics of function invocations such as the container size, the invocation frequency and recency.
Compared to LRU and FC, pCache reduces the system cost by up to 57.2$\%$ and 62.1$\%$, respectively.
The results imply that pCache maintains high performance when coping with dynamic serverless invocations.

\textbf{Distributions of system cost:}

Figure~\ref{fig: boxplot-ns-0.5} to \ref{fig: boxplot-ns-1.5} illustrate the distributions of system cost.
The box plots show the maximum, median and minimum of the results when $\beta$ ranges from 0.5 to 1.5.
In figure~\ref{fig: boxplot-ns-0.5}, the maximum system cost of pCache is 16.6 while that of LRU and FC is 125.6.
Similarly, in figure~\ref{fig: boxplot-ns-1}, we observe that pCache achieves 34.8 in the system cost while that of LRU and FC are 59.7 and 125.6, respectively.
Figure~\ref{fig: boxplot-ns-1.5} presents the distributions of system cost for $\beta = 1.5$, pCache reduces the maximum system cost by 57.2$\%$ compared to LRU and FC.
The rationale is that pCache incorporates the unique features of serverless invocations such as the invocation frequency, recency and etc so that popular containers are more likely to be cached.

\begin{figure}[tbhp]
    \centering
    \subfloat{\includegraphics[width=0.36\textwidth]{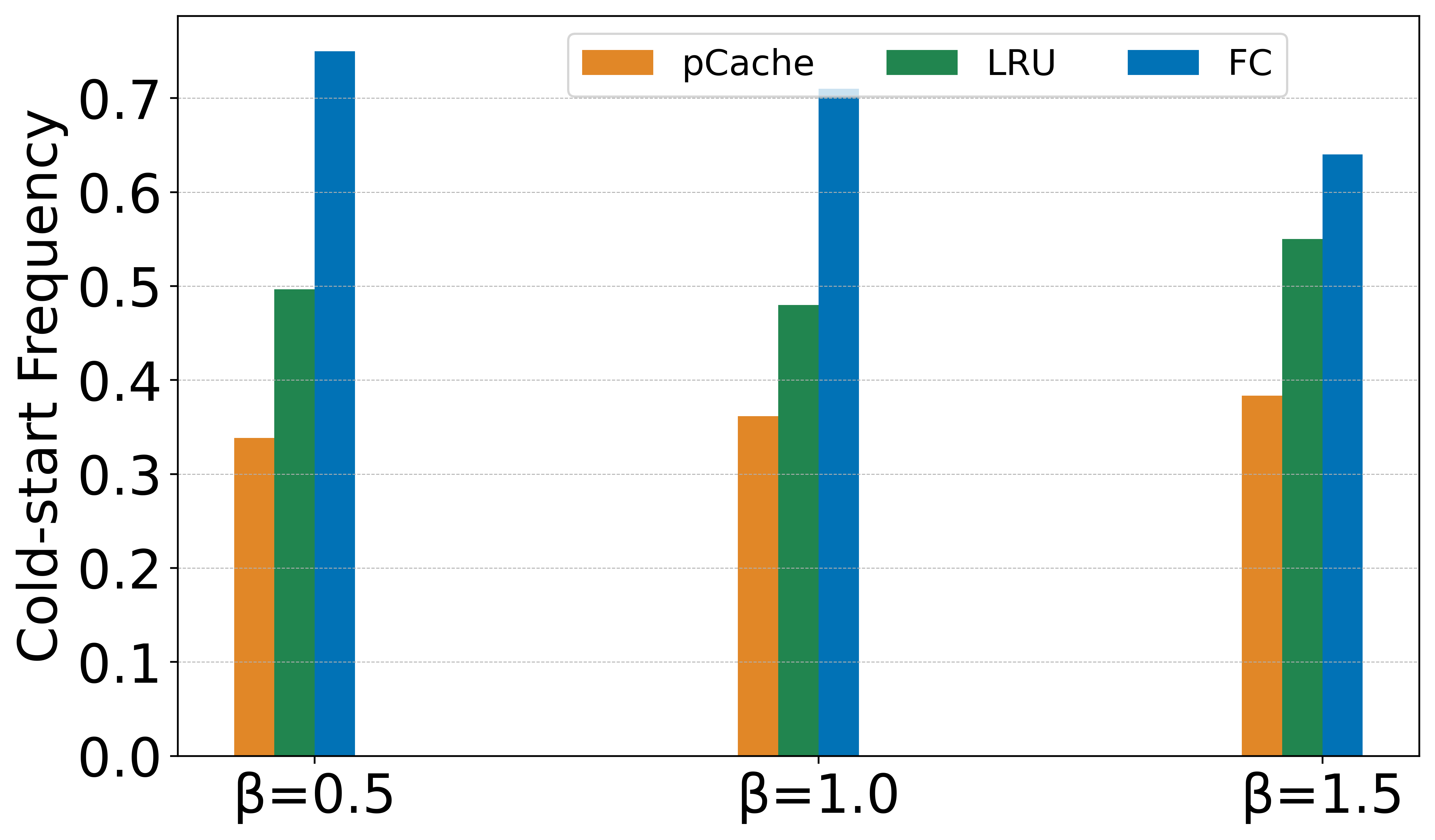}}
    \caption{Cold-start frequency in Knative}
    \label{fig: cold-k8s}
\end{figure}

\textbf{Cold-start frequency:}
Next, we evaluate the impact on cold-start frequency. The cold-start frequency refers to the frequency of cold-start happens across all invocations in the experiment.
As shown in Fig.~\ref{fig: cold-ns3}, pCache achieves the best performance in terms of cold-start frequency which are 0.224, 0.287, 0.121, respectively. 
When $\beta$ ranges from 0.5 to 1.5, LRU achieves 0.558, 0.498 and 0.19, respectively.
FC performs the worst due to the rigid caching policy.
FC caches each container for a fixed period of time which cannot adapt to burst and concurrent workloads.
LRU performs better than FC because LRU considers the recency of invocations which helps to adapt to time-varying request patterns.

\subsection{Implementation in Knative}

Knative is an open-source solution for serverless applications over Kubernetes.
In Knative, we devise a scheduler to implement the proposed algorithms.
Also, we implement the logic of request distribution into the scheduler and dynamically terminate the containers.
We implement Knative over 11 virtual machines with 4GB memory.
We reduce the number of requests in the dataset by 10000x to adapt to the system capacity.

\textbf{Performance summary in Knative} pCache achieves the best performance in terms of average system cost.
pCahce reduces the system cost by up to $25\%$ and $55.8\%$ compared to LRU and FC, respectively.
Also, we observe that pCache reduces the cold-start frequency by $32\%$ and $51.5\%$ compared to LRU and FC, respectively. 

\textbf{The average system cost:}
As illustrated in Figure~\ref{fig: average-cost-k8s}, when $\beta$ increases from 0.5 to 1.5, we observe that pCache always outperforms LRU and FC.
The rationale is that pCache assigns high eviction probability to functions that are less likely to be invoked in the near future.
Therefore, pCache is more likely to cache popular functions and hence achieves lower cold-start frequency.
With lower cold-start frequency, the system cost is effectively reduced.
Also, the results imply that pCache shows stable performance over different request distributions when $\beta$ increases. 

In Figure~\ref{fig: average-cost-k8s}, we also observe that when $\alpha$ increases, pCache always shows the best performance.
$\alpha$ is a weight parameter that impacts the weight of service latency cost and running cost.
When $\alpha$ is very small, e.g., 0.001, we will have the service latency cost significantly dominate the system cost. 
Hence, in this experiment, we increase the value of $\alpha$ from 0.001 to 0.04 to verify the performance of pCache.
The experimental results show that pCache effectively reduces the system cost by up to $55.8\%$ compared to the benchmarks over different values of $\alpha$.

\textbf{Distributions of system cost:}
Figure~\ref{fig: boxplot-knative} shows the distributions of system costs in Knative.
In Figure~\ref{fig: boxplot-k-0.5}, pCache outperforms LRU by 42.4$\%$ in light of minimum system cost.
Compared to FC, pCache reduces the median system cost by 44.9$\%$.
We observe similar trends in Figure~\ref{fig: boxplot-k-1} and Figure~\ref{fig: boxplot-k-1.5}, pCache reduces the minimum system cost by up to 42.4$\%$, the median system cost by up to 64.9$\%$, the maximum system cost by up to 52.1$\%$, respectively.
The results in Knative imply that pCache efficiently reduces the system cost in a real-world serverless platform and dynamically adapts to time-varying workloads.

\textbf{Cold-start frequency:}
Furthermore, Fig.~\ref{fig: cold-k8s} shows the cold-start frequency obtained by pCache, LRU and FC. 
pCache still achieves the lowest cold-start frequency followed by LRU and FC.  
The cold-start frequency of pCache ranges from 0.34 to 0.38 while that of LRU and FC rockets to 0.48 and 0.75, respectively. 
This is because pCache jointly considers several characteristics of serverless invocations such as the resource footprint, invocation frequency and etc which helps to avoid frequent container creation and termination.

\section{Conclusion and Future Work}\label{section:conclusion}

In this paper, we studied the online orchestration of serverless functions in edge computing with the aim of minimizing the system cost incurred by request distribution and function caching.
We prove this problem is NP-hard by analysis.
We propose an online competitive algorithm with performance guarantee which jointly considers network topology and resource constraints.
Through extensive evaluations based on trace-driven simulations and implementation over Knative, we show that the proposed algorithms outperform the baselines from state-of-the-art by up to 62.1$\%$ in system cost and up to 69.1$\%$ in cold-start frequency.

Our work also raises several open questions that are worth investigating in the future.
First, we would like to consider the serverless request distribution problem in mobile edge computing and investigate the impact of mobility.
Besides the expected challenges incurred by mobile networks, especially with the bandwidth constraints in wireless communication, we intend to further investigate how to extend our framework to more dynamic settings with joint placement and service migration.
Second, while we assume that edge nodes never fail, it is worth investigating how to migrate services when an edge node is not available or reliable.
In particular, edge servers are not as reliable as those in cloud data centers, making serverless services more vulnerable to failures.
Although deploying backup containers seems to be a promising solution to cope with such failures, it dramatically increases the system cost, which violates the spirit of serverless computing, namely pay-as-you-go.
Hence, finding the sweet spot between system cost and reliability is pivotal for efficient and reliable serverless platforms.
Our proposed framework can be modified to adapt to the extension: we consider keeping functions alive after serving requests, if we further incorporate a failure model, we can then calculate the cost of failures and the cost of provisioning backup containers, aiming to optimize the system cost under reliability constraints.

\bibliographystyle{unsrt}
{
\bibliography{serviceFunctionChain.bib}}

\end{document}